\documentclass[a4paper,11pt]{article}

\pdfoutput=1 

\usepackage{jheppub} 

\usepackage{bm}
\usepackage{color}
\usepackage{appendix}
\usepackage{graphicx}
\usepackage{amsmath}
\usepackage{amssymb}
\usepackage{slashed}
\usepackage{tabularx}
\usepackage{wrapfig}
\usepackage{slashed}

\title{ Inclusive and exclusive neutrino--nucleus cross sections and the reconstruction of the 
interaction kinematics}

\abstract {
We present  a full kinematic analysis of neutrino-nucleus charged current quasielastic interactions based on  the Local Fermi Gas model and the Random Phase Approximation. The model was implemented in the NEUT Monte  Carlo framework, which allows us to investigate potentially measurable observables, including hadron distributions. We compare the predictions simultaneously to the most recent T2K and MINERvA charged current (CC) inclusive, CC0$\pi$  and transverse kinematic-imbalance variable results. We pursuit a microscopic interpretation of the relevant reaction mechanisms, with the aim to achieving in neutrino oscillation experiments a correct reconstruction of the incoming neutrino kinematics, free of conceptual biasses. Such study is of the utmost importance for the ambitious experimental program which is underway to precisely determine neutrino properties, test the three-generation paradigm, establish the order of mass eigenstates and investigate leptonic CP violation. }

\author[a]{B.Bourguille}
\author[b]{J.Nieves}
\author[c]{F.S\'anchez\footnote{Corresponding author.}}

\affiliation[a]{Institut de Fisica d'Altes Energies (IFAE), The Barcelona Institute of Science and Technology. Edifici Cn, Universitat Autonoma de Barcelona, Bellaterra (Barcelona), Spain}

\affiliation[b]{Instituto de Fisica Corpuscular (IFIC), Centro Mixto CSIC-Universidad de Valencia, Institutos de Investigacion de Paterna, Apartado 22085, E-46071 Valencia, Spain}

\affiliation[c]{Universit\'{e} de Gen\`{e}ve - Facult\'{e} des Sciences, D\'{e}partement de Physique Nucl\'{e}aire et Corpusculaire (DPNC)
24, Quai Ernest-Ansermet, CH-1211 Gen\`{e}ve 4, Switzerland}

\emailAdd{bruno.bourguille@free.fr}
\emailAdd{jmnieves@ific.uv.es}
\emailAdd{federico.sancheznieto@unige.ch}

\begin{document}

\maketitle

\flushbottom

\section{Introduction}

The studies of neutrino-nucleus interactions are entering a new stage, motivated by long-baseline experimental programs, in which the statistical uncertainties will diminish and thus the nuclear effects -- contributing to the systematical error -- have to be kept well under control~\cite{Alvarez-Ruso:2017oui}. The incomplete theoretical knowledge of the neutrino-nucleus interactions influences various stages of experimental analysis.
For instance, for the future Hyper-Kamiokande water Cherenkov detector~\cite{Abe:2014oxa}, the method for reconstructing the neutrino energy will be mainly based on the kinematics of the outgoing muon, which is the only particle observed, assuming that the reaction-mechanism is  two-body charged-current quasielastic (CCQE) dispersion on a bound nucleon. However, the energy range of the neutrino flux produced in the J-PARC facility~\cite{Abe:2012av}, extending beyond 10 GeV, is such that other physical mechanisms give non negligible contributions to the cross section. In particular,  multi-nucleon knockout processes (mainly driven by the excitation of two particle-two hole, 2p2h, components in nuclei) should be taken into account. Since in the latter processes the interaction takes place on a pair of nucleons, the energy balance is different than in the QE case, driven by the excitation of only one nucleon (1p1h). Mismatching the signal coming from these two reaction mechanisms would lead to a bias in the energy reconstruction~\cite{Nieves:2012yz,Alvarez-Ruso:2014bla}. It is therefore crucial to properly include the 2p2h channel into the Monte Carlo (MC) event generators.

\begin{figure}
\begin{center}
\includegraphics[width=0.47\textwidth]{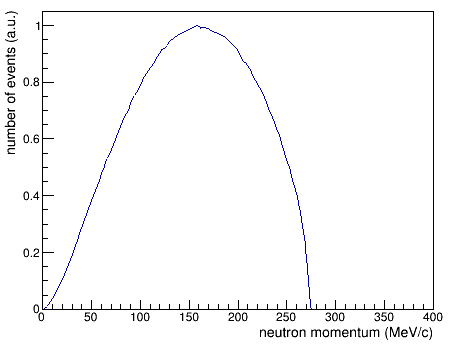}
\includegraphics[width=0.52\textwidth]{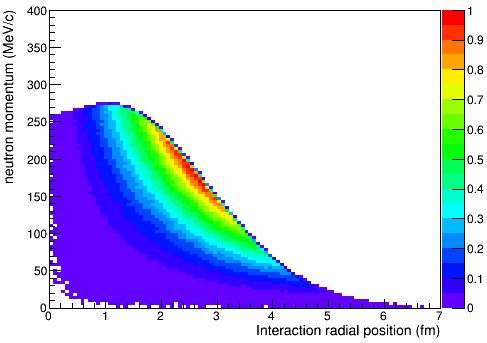}
\caption{\label{Fig:NeutronMomentum} Left: LFG neutron momentum distribution in carbon. Right: Two-dimensional neutron momentum distribution  correlated to the radial interaction point inside the nucleus. Distributions are folded with the T2K $\nu-$energy flux \cite{Abe:2012av}. }
\end{center}
\end{figure}

\section{ CCQE $\rm 1p1h$ model } 

We developed a full exclusive CCQE 1p1h MC event generator based on the theoretical scheme developed in \cite{Nieves:2004wx}. The model is capable to simulate both neutrinos ($\nu_{\mu} n \to \mu^{-} p$) and anti-neutrino\footnote{In general, we will refer to neutrino-induced reactions, unless it is  explicitly mentioned that the discussion is about processes with anti-neutrinos. } ($\bar\nu_{\mu} p \to \mu^{+} n$) reactions for a  variety of nuclei: C,O, Al, Ti, Fe and Ca. The original code \cite{Nieves:2004wx, Nieves:2011pp} provided the total cross-section for a fixed neutrino energy value, and the CCQE differential cross sections depending on the energy and solid angle of the outgoing charged lepton. 
The code was included in the NEUT MC generator \cite{Hayato:2009zz}. The implemented modifications in NEUT  keep all the physics of the model: Local Fermi Gas (LFG) nucleon-dynamics, short and long range Random Phase Approximation (RPA) correlations, Pauli blocking, lepton Coulomb corrections,... and provide an almost fully exclusive cross-section by predicting the hadron kinematics in the first step (weak absorption of the gauge boson) of the reaction.

\subsection{ Implementation of the exclusive CCQE model in the MC}  

We implement the QE model of Ref.~\cite{Nieves:2004wx} in NEUT MC, so that we could extract both the position of the first step interaction and full hadron kinematics. The model is almost fully exclusive since we compute the cross-section as  function of the: 

\begin{itemize} 

\item radial position of the interaction in the nucleus and modulus 
of the target (hit) nucleon three-momentum from the LFG distribution.

\item (anti-)neutrino energy. 

\item outgoing lepton momentum and angle. 

\item angle between the outgoing proton (neutron) and the transfer momentum direction (we test for Pauli blocking at the given radial position).

\item The azimuthal angle of the final hadron with respect  to the lepton reaction plane. This angle is generated with  a flat probability in the   $[-\pi,\pi]$ interval

\end{itemize}
With this information we can obtain the whole event kinematics applying conservation of momentum and energy:

\begin{itemize} 

\item lepton four-momentum $p^{\mu}_\ell$.

\item  target (hit) nucleon four-momentum $p^{\mu}_N$.

\item final state nucleon four-momentum $p^{\mu}_{N'}$.

\end{itemize}

\begin{figure}
\begin{center}
\includegraphics[width=5.cm]{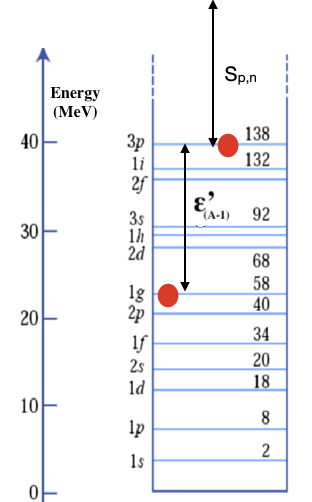}
\caption{\label{Fig:Excitation}Pictorial representation of the excitation energy scheme. When a nucleon is removed from a deep shell, the hole energy  remains until the nucleus is de-excited. The FG removal energy refers  only to Fermi level nucleons, which would correspond to leave the daughter $(A-1)$ nucleus in the ground state within this statistical nuclear model. It is an approximation to the experimental proton and neutron separation energies ($S_{p,n}$) of the target nucleus.  Knocking out a nucleon from deeper levels will give 
rise to excited final-nucleus states,  equivalent to the binding energy of the nucleon occupying this level ($\epsilon$). We also show the accumulative occupation numbers when additional shells are considered. The energy scale depicted at the left of the plot has an arbitrary origin,  and it is only intended to illustrate the energy differences between shells. }
\end{center}
\end{figure}

\subsection{Local Fermi Gas and nucleon kinematics: implementation of the removal energy correction}

The present model utilizes a LFG to describe the nucleus, which provides on one hand a more accurate description of the Fermi momentum and Pauli blocking than those obtained in global FG approaches. On the other hand, it allows to locate the 
position of the first interaction inside the nucleus, which might affect the strength/relevance of the nuclear re-interactions. We will discuss interactions of neutrinos off carbon, which is the main target material for the most recent neutrino scattering experiments: NOvA, T2K, MINERvA and MiniBooNE. Figure \ref{Fig:NeutronMomentum} shows that in this nucleus, the interactions mostly occur between 1.5 and 4~fm. The LFG model introduces a relation between the  Fermi momentum and the radial position given by the equation: 
\begin{equation}
p_{Fn}=(3 \pi^{2}\rho_{n}(r))^{1/3} \label{eq:pF}
\end{equation}
with $\rho_{n}(r)$, the density of neutrons (protons for anti-neutrino reactions) for a given radial position, $r$, inside the nucleus.
In our MC,  we had chosen the neutron (hit nucleon)  momentum to be taken between 0 and this local Fermi momentum. The neutron momentum as function of the radial position of the interaction is shown in Fig.~\ref{Fig:NeutronMomentum}. We can see that the highest local Fermi momentum is achieved at radius slightly above 1~fm for carbon.

\begin{figure}
\begin{center}
\includegraphics[width=0.325\textwidth]{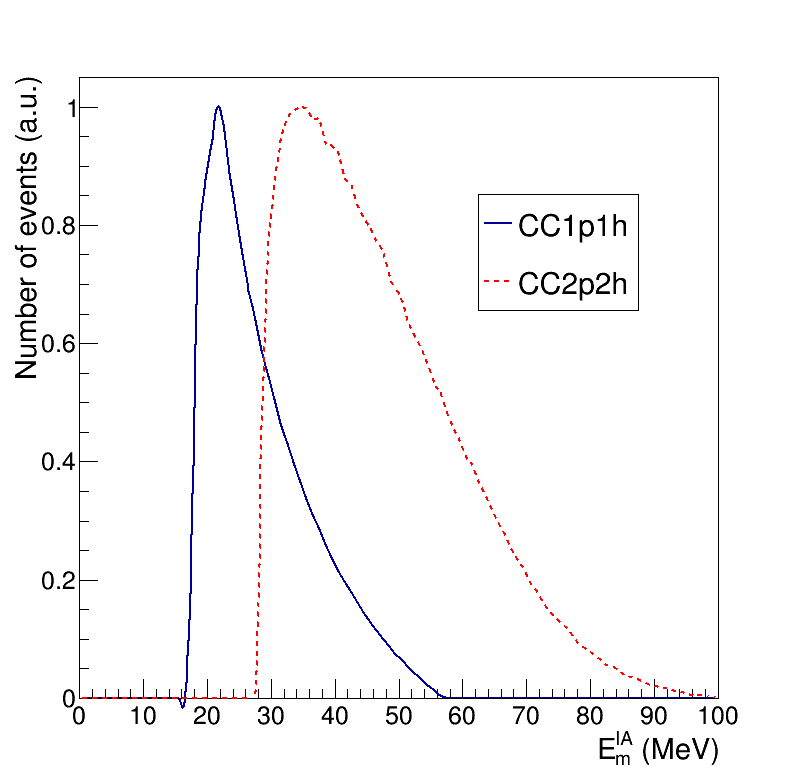}
\includegraphics[width=0.325\textwidth]{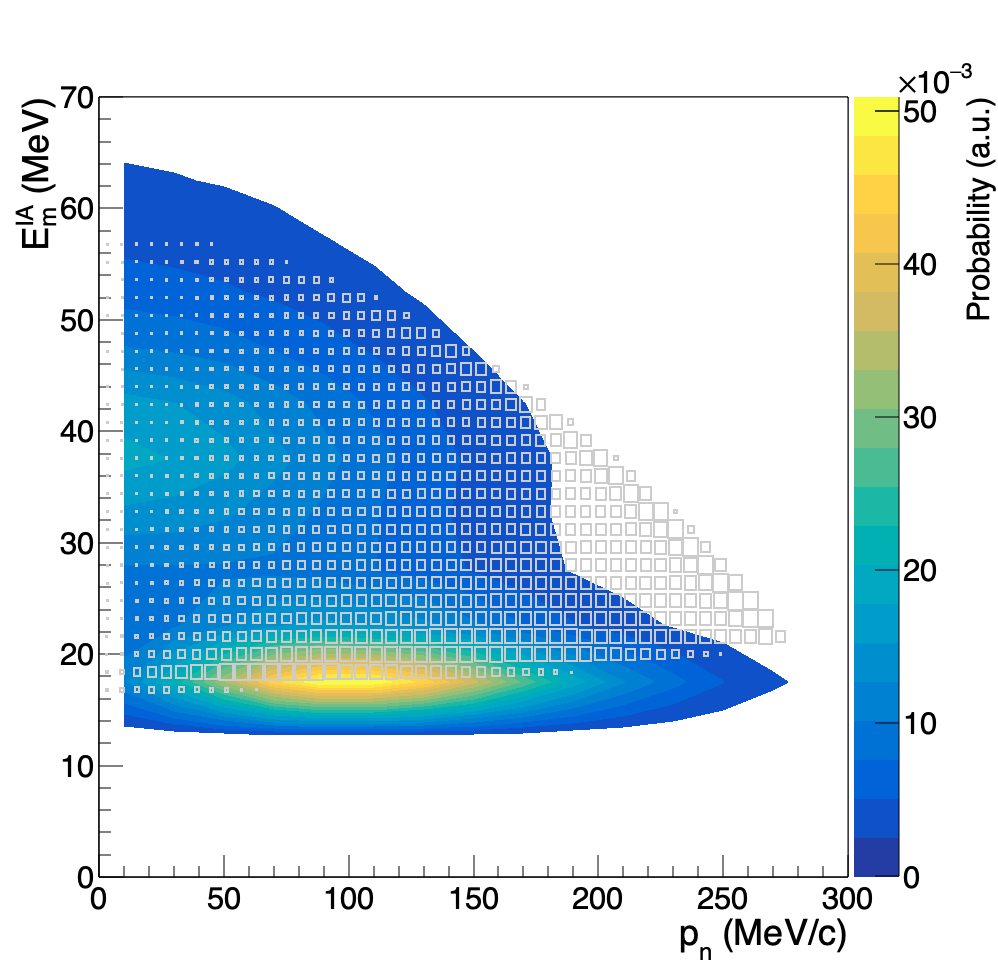}
\includegraphics[width=0.325\textwidth]{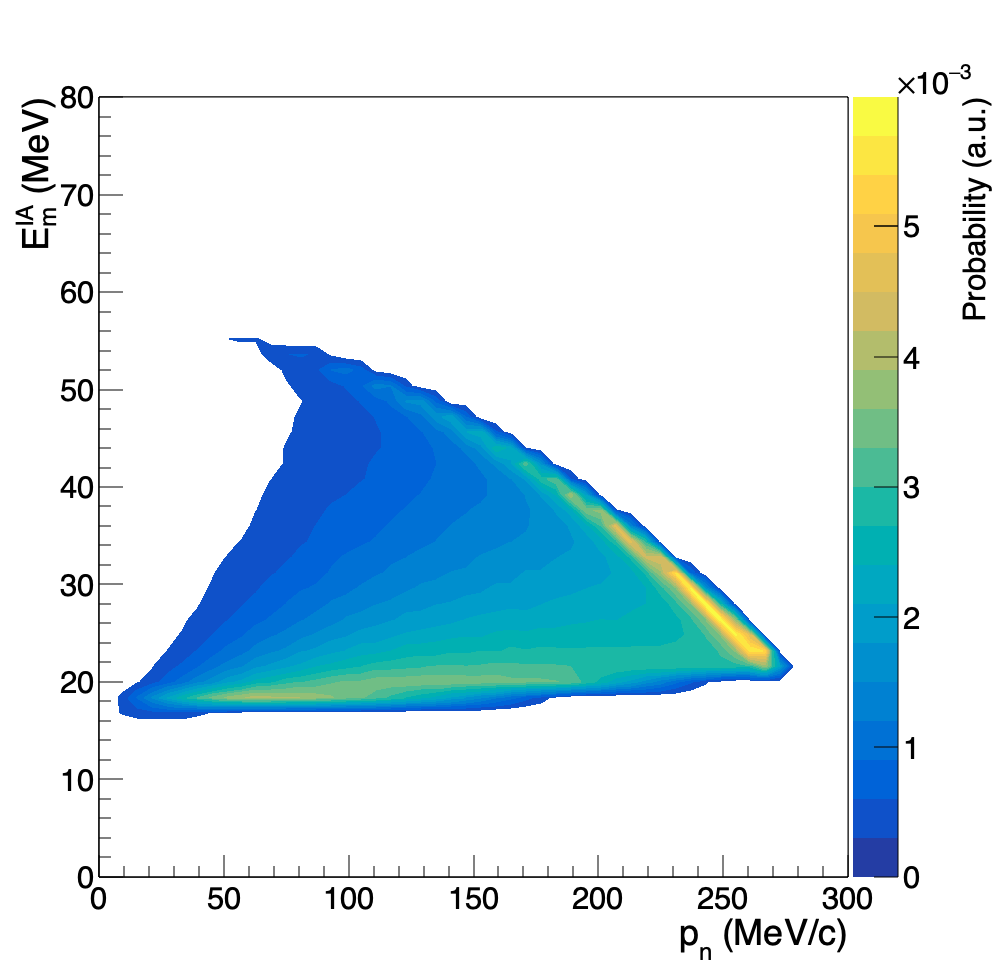}

\caption{\label{Fig:TFermi} Left: Number of events as function of the $E_{m}^{IA}$ energy for neutrino scattering off $^{12}$C within the LFG model. Both the 1p1h (Eq.~\eqref{eq:Emiss}) and 2p2h (Eq.~\eqref{eq:Emiss-2p2h}) $E_{m}^{IA}$ distributions are shown by the blue-solid and red-dashed lines, respectively. The gap between the two distributions is caused by the excitation energy of the two holes in the final state. As in Fig.~\ref{Fig:NeutronMomentum},  results have been folded with the T2K neutrino energy flux. Center: Probability to find a neutron in carbon with a momentum ($p_n$) for a given reaction missing energy ($E_m^{IA}$) (see Eq. \eqref{eq:Emiss}) as predicted by the SF model~\cite{Benhar:1989aw, Benhar:1994hw, Benhar:2005dj} (contour plot) and for this implementation of the LFG (box plot). Right: LFG predictions corresponding to the box plot displayed in the middle panel. In all cases, the T2K flux~\cite{Abe:2012av} is used.  }
\end{center}
\end{figure}

At first, the  model relies on the Impulse Approximation (IA), where we consider the hit-nucleon is a plane wave state, and the momentum balance reads
\begin{equation}
    \vec{p}_\nu + \vec{p}_{N} + \vec{p}_{A-1}= \vec {p}_{\mu} + \vec{p}_{N'} + \vec{p}^{\,\prime}_{A-1} \label{eq:mom_con}
\end{equation}
where $\vec{p}_{A-1}$ is the momentum of the remaining $(A-1)-$nucleons at the moment of  the collision. This momentum should cancel with the hit-nucleon momentum ($\vec{p}_N$) so the total momentum of the initial nucleus vanishes. Within this approximation, we also consider that the momentum of the final state nucleus ($\vec{p}^{\,\prime}_{A-1}$) is equal to the residual momentum of the initial nucleus and both cancel out  ($\vec{p}^{\,\prime}_{A-1}= \vec{p}_{A-1}$). Thus the balance of Eq.~\eqref{eq:mom_con} reduces to: 
\begin{equation}
    \vec{p}_\nu + \vec{p}_N = \vec{p}_\mu + \vec{p}_{N'} \label{eq:balmon}
\end{equation}
which is the traditional equation of momentum conservation within the IA model.  On the contrary, the IA is broken for the energy balance, due to the need of an energy contribution to transit from the ground state of the target nucleus to a new final nuclear configuration, with the daughter nucleus left in its ground or an excited state or even broken. Actually, the energy conservation equation reads
\begin{equation}
E_\nu + M_A = E_\mu + E^{\infty}_{N'} + M^\prime_{A-1} + \epsilon^\prime_{A-1} + T^\prime_{A-1} 
\label{eq:ene_con}
\end{equation}
where $M_A$ and $M^\prime_{A-1}$ are the the ground state masses of the initial and final nuclei, and $E^{\infty}_{N'}$ is the energy of the asymptotically observed nucleon ($N'$). In addition, $\epsilon^\prime_{A-1} > 0$ is the excitation energy of the final nucleus, which average is expected to be between 10 and 20~MeV (see Fig.~\ref{Fig:Excitation}). Finally,  $T^\prime_{A-1}$ is the final nucleus kinetic energy, which  is very small (typically of the order of $p_{F_n}^2/(2M_{A-1})\sim 2$ MeV for carbon target)  and it is  approximated to zero in what follows.

The energy balance in Eq.~\eqref{eq:ene_con} does not apply to cases where any secondary re-scattering collision changes the energy of the nucleon that emerges after the weak absorption of the gauge boson\footnote{In fact, in these latter situations, the rupture of the daughter nucleus might occur and the analysis is more complicated.}.

The excitation energy can be estimated, in a first approximation, to be the energy of the hole, within the FG model\footnote{Any LFG model implicitly assumes the existence of a mean-field potential $U=-T_F$, which cancels in the difference of energies, and binds the nucleons.  }

\begin{figure}
\begin{center}
\includegraphics[width=0.47\textwidth]{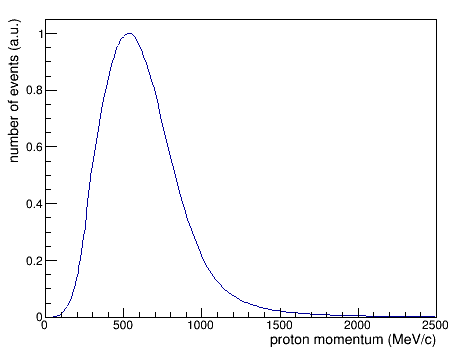}
\includegraphics[width=0.50\textwidth]{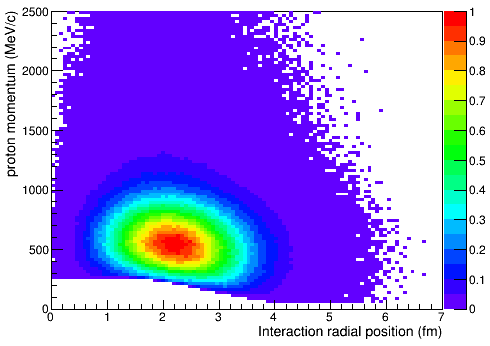}

\caption{\label{Fig:ProtonMomentum} Left: Primary proton,  created after the absorption of the gauge boson $W^+$ in neutrino processes,  momentum distribution in carbon predicted by the model. Right: Proton momentum distribution predicted by the model as function of the radial interaction point inside the nucleus. As in Fig.~\ref{Fig:NeutronMomentum},  results have been folded with the T2K neutrino energy flux.}
\end{center}
\end{figure}

\begin{equation}
\epsilon^\prime_{A-1} \sim (T_F - T_{N} )
\end{equation}
with $T_F$ the kinetic energy of the nucleon at the Fermi level for the given radial position, and $T_N$ the actual kinetic energy of the knocked out nucleon in the target nucleus. On the other hand, the experimental  nucleon separation energy $S_{N}$  can be obtained from the masses of the initial and final nuclei: 
\begin{equation}
 S_{N} =  M^\prime_{A-1}-M_{A}+m_{N}
\end{equation}
where  $M_{A}$ and $ M^\prime_{A-1}$ are the ground-state  masses of the initial ($A_Z$) and final [$(A-1)_Z$  or $(A-1)_{Z-1}$ for neutrino or anti-neutrino reactions]  nuclei, and $m_{N}$ the mass of the target nucleon. Re-writing Eq.~\eqref{eq:ene_con} using $S_{N}$, we obtain:
\begin{equation}
E_{\nu} + \left(m_N+T_{N}-T_F\right) = E_{\mu}+S_{N}   + E^{\infty}_{N'} \label{eq:ene-bal}
\end{equation}
which reduces to  the usual IA  energy conservation formula, but with an additional correction: the term $S_{N} + T_F $, which is an approximation of the experimental removal energy and takes into account to  some level  the excitation of the final state nucleus. 
 In case of the Relativistic Global Fermi Gas (RGFG), $T_F$ takes a constant value $(\sim 27$ MeV), but in the 
relativistic LFG model, it has a distribution depending on the radial position of the interaction. This dependency introduces several removal energies simulating a continuous distribution of excitation of nuclear states.  

\begin{figure}
\begin{center}
\includegraphics[width=0.495\textwidth]{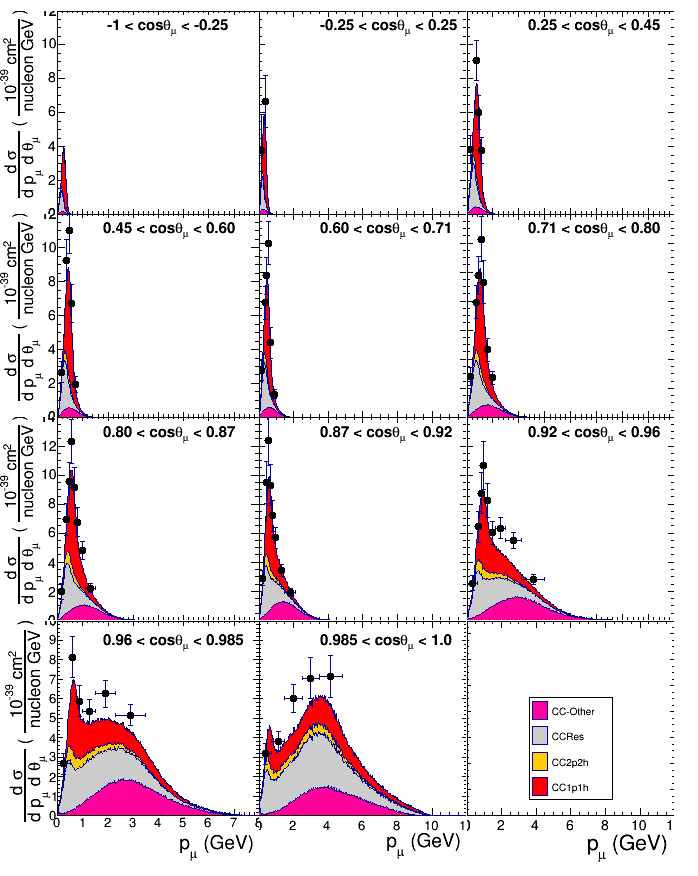}
\includegraphics[width=0.48\textwidth]{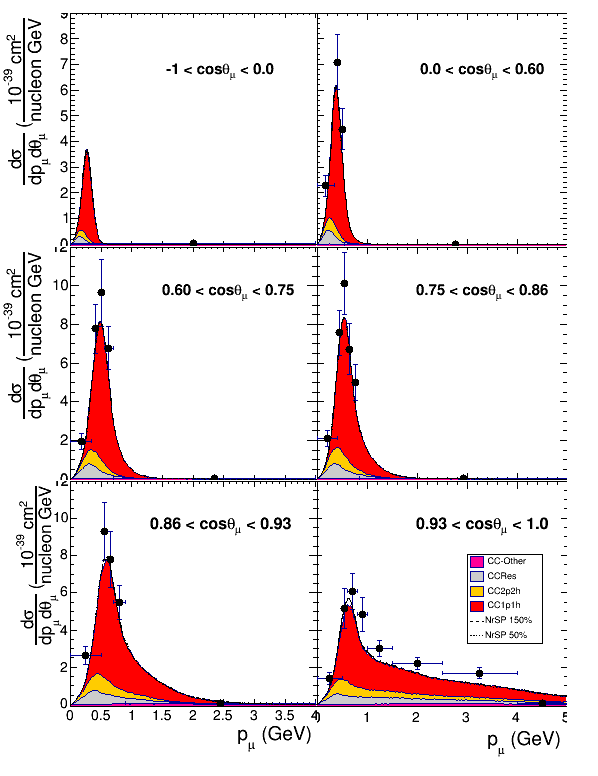}
\caption{ T2K CC (left) and CC0$\pi$ (right) inclusive double differential cross-section $d\sigma / d p_{\mu} d \cos {\theta_{\mu}}$. Data, taken from Refs.~\cite{Abe:2018uhf} and \cite{Abe:2020uub}, respectively, are compared to the results obtained from the present implementation of NEUT with the numerical values for the $\chi^2-$likelihood test compiled in Table~\ref{Tab:T2KChi2}. Last bin accumulates all the statistics until the kinematic limit of 30~GeV.\label{Fig:CCincT2K}}
\end{center}
\end{figure}

This discussion  of the IA energy balance  fixes a problem with the nuclear missing energy ($E_m^{IA}$), which appears within the traditional implementation  in MC event generators of the relativistic LFG and RGFG models
\begin{equation}
E_{m}^{IA} = E_{\nu}-E_{\mu}-T^{\infty}_{N'} \label{eq:Emiss}
= S_{N}+(m_{N'}-m_N) + T_F -T_N
\end{equation}
with $E^{\infty}_{N'}= T^{\infty}_{N'}+ m_{N'}$. The value of $E_m^{IA}$  becomes negative (non-physical) for some  values of $T_{N}$ when, as it is common, the $T_F$ correction is not added\footnote{This is taken into account in some models such as NuWro by adding a constant that  restores the validity of the model. }. Equivalently, the problem is caused by the wrong assumption of taking $E_N= \left(m_N+T_{N}\right)$, instead of the correct expression $E_N= \left(m_N+T_{N}-T_F\right)$, which includes the mean field potential responsible for binding the nuclear system. The distribution $E_{m}^{IA}$ of energies for a relativistic LFG  is depicted in the left plot of Fig.~\ref{Fig:TFermi} for neutrino scattering off carbon, where $S_{n}+(m_{p}-m_n)= \Delta(^{11}{\rm C})-\Delta(^{12}{\rm C})+\Delta(^{1}{\rm H})= 17.4$ MeV [$\Delta(A_Z)$ is the mass excess of  the $A_Z$ nucleus]. The average $E^{IA}_{m}$ is 28 MeV, very similar to the binding energies used in RGFG models (25~MeV)\cite{Abe:2019arf} or in MINERvA   (27.13~MeV)\cite{Lu:2018stk}. In addition, we can use this average of $E^{IA}_{m}$ for the LFG model in carbon, to estimate the average  of 
the excitation energy $\langle \epsilon^\prime_{A-1}\rangle\sim \langle T_F - T_{N} \rangle \sim 11$ MeV for this target, using $S_{n}+(m_{p}-m_n)=17.4$ MeV.

\begin{figure}
\begin{center} 
\includegraphics[width=0.655\textwidth]{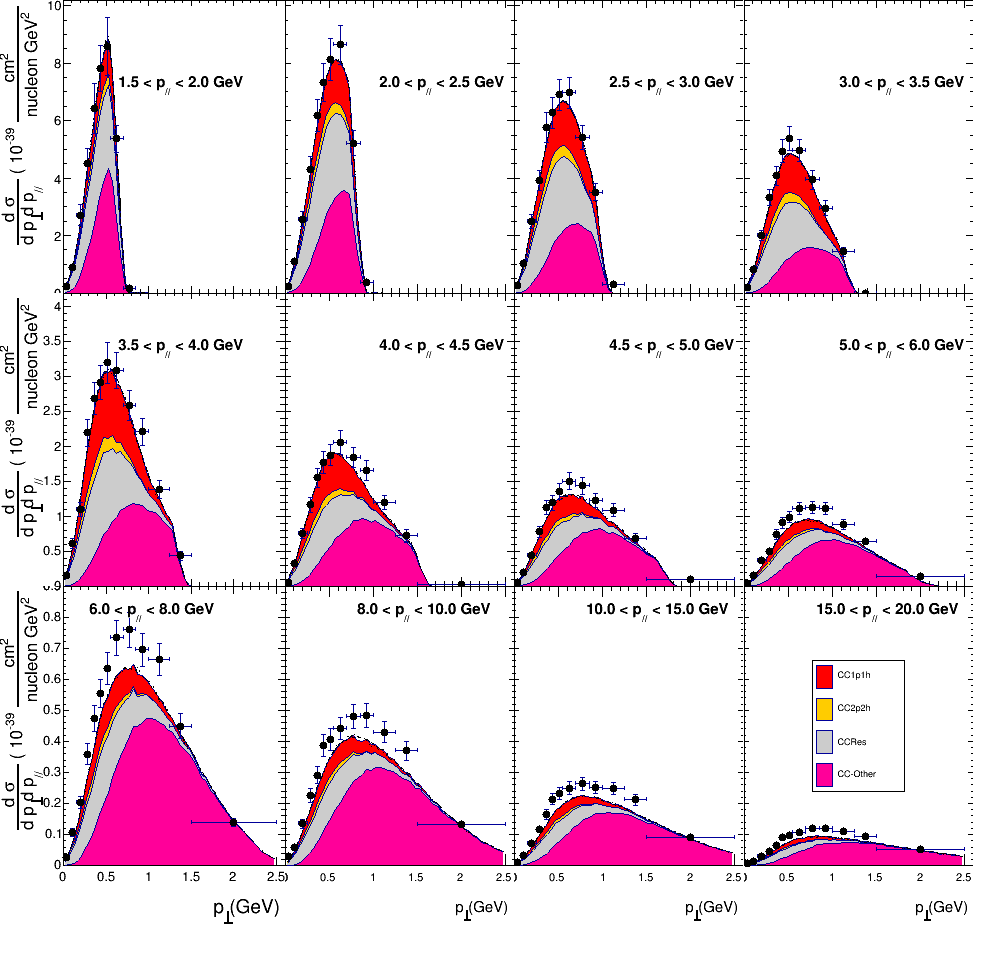}
\includegraphics[width=0.655\textwidth]{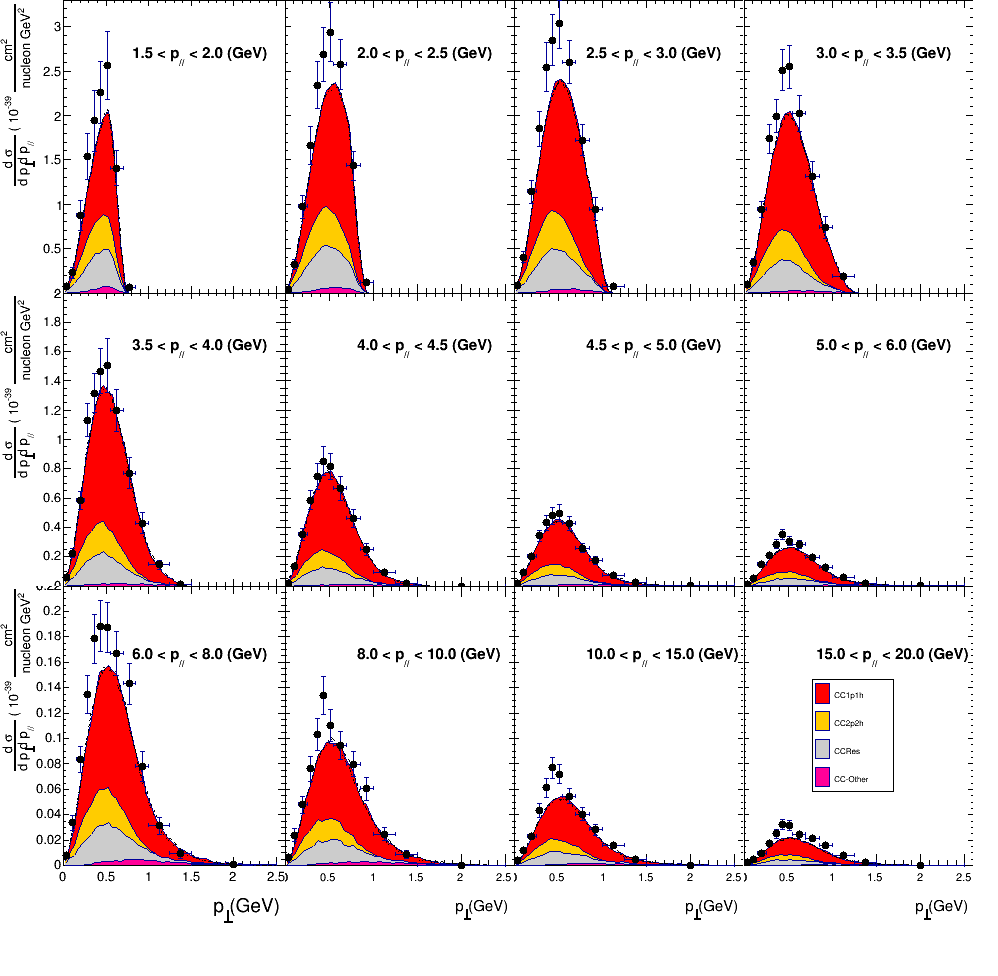}
\caption{\label{Fig:MinervaCCPT} MINERvA CC (top) and CC0$\pi$ (bottom) inclusive double differential cross-section $d^2\sigma / d p_{\perp} d p_{\parallel}$.  Data, taken from Refs.~\cite{Filkins:2020xol} and \cite{Ruterbories:2018gub}, respectively,  are compared to the results obtained from the present implementation of NEUT, with the $\chi^2-$likelihood tests compiled in Table~\ref{Tab:MinervaChi2}.}
\end{center}
\end{figure}

We pay now attention to the two-dimensional $(p_N,E_m^{IA})$ distribution shown in the middle and right plots of Fig.~\ref{Fig:TFermi} following the same representation as in the Spectral Function (SF) scheme~\cite{Benhar:1989aw, Benhar:1994hw, Benhar:2005dj}. The carbon SF obtained in \cite{Benhar:1994hw} is comprised of two contributions. The first one is determined by a mean-field description of the nucleus, while the second one  takes into account two-nucleon short range correlations, and it is computed within a correlated basis function 
scheme in isospin-symmetric nuclear matter. The mean-field contribution of the SF modifies the dispersion relation by forcing a set of effective bound masses. This way the value of $E^{IA}_m$ is constant for each of the nuclear levels with a broad momentum distribution, which is additionally distorted by the contribution of the correlated part of the SF. 
The model presented here is based on  the LFG approach to the nucleus, where the dispersion relation is fixed to the on-shell target nucleon mass\footnote{The 1p1h contribution to the nuclear response function depends on the energy difference between particle- and hole- nucleons, where the mean-field potential $-T_F(r)$ cancels out.}, but with a Fermi level that depends on the spatial position through the local density. Despite its simplicity, the LFG distribution, as shown in Fig.~\ref{Fig:TFermi}, follows a pattern  similar to that exhibited by the more realistic one inferred  from the SF scheme of Refs.~\cite{Benhar:1989aw, Benhar:1994hw, Benhar:2005dj}. Nevertheless, some differences between both sets of predictions are visible in Fig.~\ref{Fig:TFermi}, in particular at the edges of the $(p_N,E_m^{IA})$ contours.   This different dependencies might introduce distinctive differences in the nuclear response when the nucleon target momentum is relevant such as in the case of low energy neutrino interactions and it might explain some of the disagreements discussed later in this work. 

\subsection{ Predictions of the final state hadron kinematics }

In the left panel of Fig.~\ref{Fig:ProtonMomentum} we show the momentum distribution of the primary proton,  created after the absorption of the gauge boson in neutrino processes, as predicted by the model presented in this work. In the right plot of the figure, we show the proton momentum correlated to the radial position of the primary interaction. Low energy protons are produced close to the outer surface of the nucleus having a large probability to survive nuclear re-scattering.  The maximum momentum of the proton is limited by the energy of the neutrino, but the lowest values are determined by Pauli blocking, which  is also function of the radial position of the interaction. The fact that the Pauli effects become less relevant  at large radii (4~fm) allows the proton momentum to have values close to zero contrary to less sophisticated models such as the global FG~\cite{Hayato:2009zz}.  

\section{Consistent implementation of the CC 2p2h model and the secondary nuclear collisions }

In NEUT~\cite{Hayato:2009zz}, the events are generated according to the distribution of the outgoing lepton, i.e. using the weight given by the value of the double-differential inclusive cross section, expressed as the contraction of lepton  and hadron tensors, as given for instance in   Eq.~(2) of Ref.~\cite{Nieves:2011pp}. In addition to the 1p1h term, the hadron tensor $W^{\mu\nu}$  accounts also for 2p2h 
contributions evaluated following the LFG scheme of Ref.~\cite{Nieves:2011pp}, which is fully consistent with the 1p1h implementation outlined above and based on \cite{Nieves:2004wx}. It is computed, for (anti)neutrino reactions,  separately for proton-neutron and proton-proton (neutron-neutron)  final states to provide isospin dependent final states.  The location of the interaction vertex in the nucleus is chosen according to the density profile, and the initial state nucleons are picked below the Fermi level corresponding to the radial position following the LFG model recipe.  
 The outgoing nucleons at the weak vertex are distributed according to the  available phase-space. This is because all the hadron-variables are integrated out in the calculation of the inclusive lepton cross sections carried out in \cite{Nieves:2011pp}. At this respect, note that in the recent re-computation of Ref.~\cite{Sobczyk:2020dkn} some of these integrations have been undone, opening the possibility to improve on this phase-space prescription. The final state nucleons are generated uniformly in the center of the mass of the hadronic system and boosted to the laboratory rest frame. Next, their momenta are tested against the local Fermi level to implement Pauli blocking. This procedure neglects the dynamics of underlying nuclear model and produces a symmetric distribution of outgoing nucleons \cite{Sobczyk:2020dkn}. The produced pair of nucleons is fed into the NEUT cascade model accounting for the transport of nucleons in the high density nuclear medium.

\begin{figure}
\begin{center} 
\includegraphics[width=0.4\textwidth]{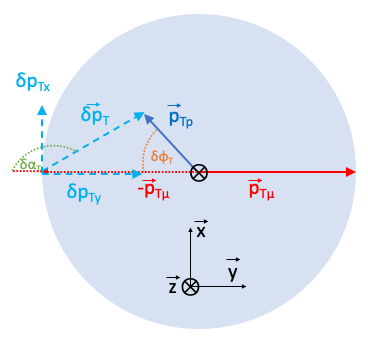}
\caption{\label{Fig:transverse} Angular ($\delta \alpha_T$,$\delta \phi_t$) and transverse momentum variables ($\vec \delta p_T$, $\delta p_{T_X}$, $\delta p_{T_Y}$)  in a reference system, where the neutrino is entering perpendicular to the plane. }
\end{center}
\end{figure}

\begin{figure}
\begin{center}
\includegraphics[width=0.495\textwidth]{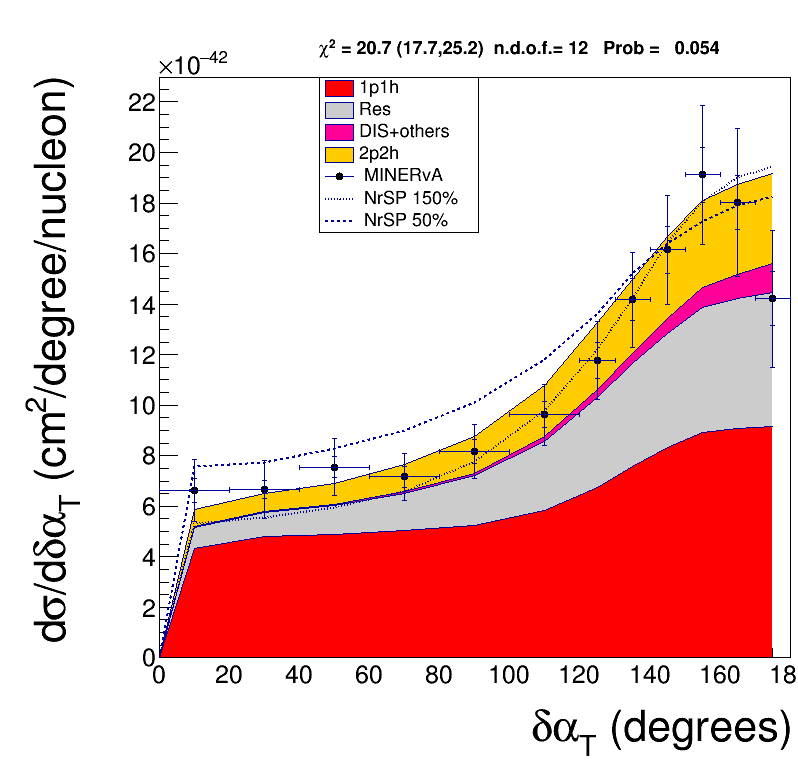}
\includegraphics[width=0.495\textwidth]{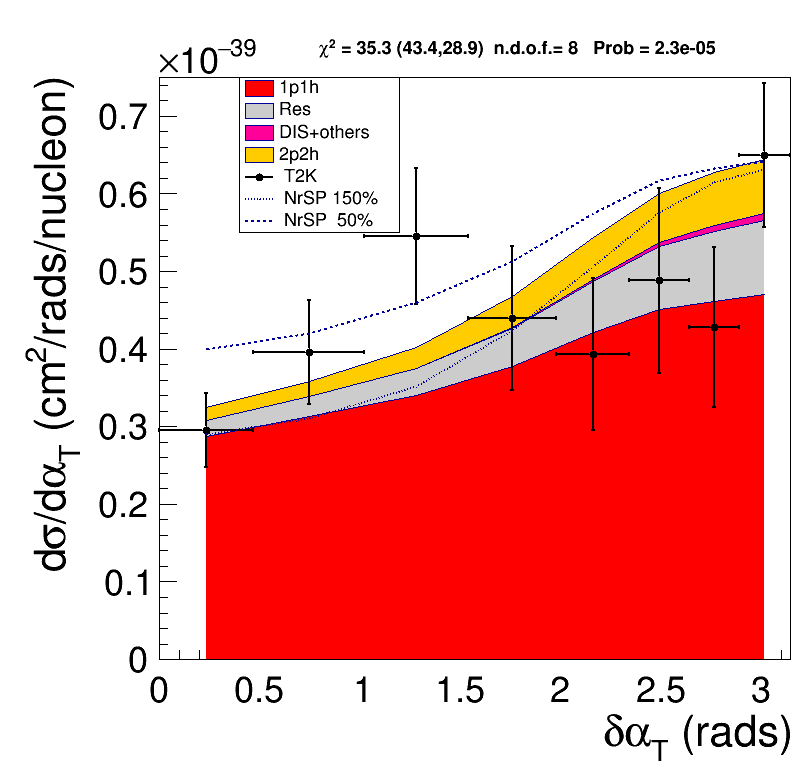}
\caption{\label{Fig:Dalphat}  CC$0\pi 1p$ MINERvA~\cite{Lu:2018stk,Cai:2019hpx,Harewood:2019rzy} (left) and T2K~\cite{Abe:2018pwo} (right) $\delta \alpha_T$ distributions compared to the predictions of the implementation of NEUT in this work. The simulation has been done for three  configurations of the NrSP (nominal, 50\% and 150\% strengths) and the obtained absolute $\chi^2-$values are compiled in Tables~\ref{Tab:T2KChi2} and \ref{Tab:MinervaChi2}.  }
\end{center}
\end{figure}

 We have introduced the same removal energy corrections than in the case of the 1p1h,  but we need to take into account the fact that two nucleons are removed  from  the nucleus. This provides an energy balance equation: 

\begin{eqnarray}
E_{\nu} &=& E_{\mu} + E^{\infty}_{N'_1}  + E^{\infty}_{N'_2} + S_{N_1N_2}  \nonumber \\
&-&\left(m_{N_1}+T_{N_1}-T_{F_1}\right)- \left(m_{N_2}+T_{N_2}-T_{F_2}\right) 
\end{eqnarray}
where we have neglected the kinetic energy of the daughter nucleus, have contemplated the possibility of different (isospin) Fermi levels for the hit nucleons $N_1$ and $N_2$, and 
\begin{equation}
   S_{N_1N_2} = M'_{A-2}-M_A+m_{N_1}+m_{N_2} 
\end{equation}
Finally, $N'_1$ and $N'_2$ are the two outgoing nucleons asymptotically observed. 
In the left plot of Fig.~\ref{Fig:TFermi}, we showed the distribution of missing energies from the 2p2h mechanism
\begin{equation}
E_{m}^{IA}\Big|_{\rm 2p2h}=E_{\nu}-E_{\mu}-T^{\infty}_{N'_1}-T^{\infty}_{N'_2}\, ,  \label{eq:Emiss-2p2h}
\end{equation}
and compared  to the 1p1h $E_{m}^{IA}$ values discussed in the previous section. The excitation of two nucleons  leads to a bigger offset of the $E_{m}^{IA}$ values and a longer energy tail  compared to that of the 1p1h distribution.  The immediate consequence of this implementation is that the average $E^{IA}_m$ is around 45~MeV which is larger than the one of the CC1p1h and larger than previous implementations of CC2p2h models, where the typical CC1p1h $E^{IA}_m$ were implemented. This correction will reduce the overall 2p2h cross-section for low energy transfers.

The remaining neutrino-nucleus interaction channels, mostly with associated pion production, are simulated based on the existing NEUT~\cite{Hayato:2009zz} Monte Carlo event generator. The resonant pion production is based on the Rein-Sehgal model~\cite{Rein:1980wg}, taking into account eighteen resonances with masses below 2~GeV and their interference terms, with the axial mass fixed to $M_\text{A}^\text{RES} = 1.21$~GeV. This model has been compared to experimental data by the T2K collaboration showing remarkable agreement\cite{Abe:2019arf}. Neutral current and charged current coherent pion production is simulated using the Rein-Sehgal model in Ref.~\cite{Rein:1982pf}. The CC coherent pion production includes  PCAC (partially conserved axial-vector current) and lepton mass corrections, as discussed in \cite{Rein:2006di}. DIS processes are simulated using the GRV98~\cite{Gluck:1998xa} parton distribution, with low-Q$^2$ corrections from the Bodek and Yang model~\cite{Bodek:2003wc}. Secondary interactions of pions inside the nucleus are simulated using an intra-nuclear cascade model based on the method developed by Salcedo et al.~\cite{Salcedo:1987md}, tuned to external $\pi-$$^{12}$C data \cite{PinzonGuerra:2018rju}.

\section{ Comparison to experimental data } 

In this section we discuss the comparison of the predictions from the  model  with  recent data from MINERvA and T2K cross sections  with no pions in the final state. The implementation of the model inside NEUT allows us to make a direct comparison with the experimental cross-sections since all the interaction channels are considered including, the transport of the nucleons and pions inside of the nucleus. The data selected include inclusive muon kinematics and TKI variables to explore the limits of the hadron kinematic predictions of the model. 

\subsection{Event Simulation and data selection }

Events are simulated using the NEUT package with the CCQE and CC2p2h reaction-mechanisms described above. We take the fluxes from the experiment releases according to their best understanding. The simulation is done for three  configurations of the nucleon  re-scattering probability (NrSP): nominal, and  50\% and 150\% strengths. This is only applied to the proton re-scattering while pions are kept to their nominal NEUT values. We select events according to the particles emitted by the nucleus after the interaction taking into account the event acceptance of the experiments as described in their published designs. 

\subsubsection{T2K data samples }

The neutrino T2K data-sample has  different selections obtained from the off-axis muon neutrino beam, which peaks around 0.6 GeV but it contains a large energy-tail ranging to the region of tens of GeV. The CC inclusive measurement considers only the muon production kinematics ignoring all hadronic activity \cite{Abe:2018uhf}. 
The T2K selection criterion for CC0$\pi$ \cite{Abe:2020uub} requires no charged or neutral pions in the final state. Based on this selection, T2K provides  double differential cross-sections for carbon and oxygen nuclear targets. The comparison is performed only on carbon data to keep a common nuclear target across the different measurements and experiments. 

The  T2K selection  criterion for the single TKI variables analysis of the CC0$\pi1p$ sample requires the detection of a muon and proton with the following conditions \cite{Abe:2018pwo}: 

\begin{itemize} 
\item  $0.45~{\rm GeV} < |\vec{p}_p|<  1.0 {\rm ~GeV}$.
\item  $ \cos{\theta_{p}} > 0.4 $ 
\item  $0.25 {\rm ~GeV} < |\vec{p}_{\mu}| < 10 {\rm ~GeV}$.
\item  $ \cos{\theta_{\mu}} > -0.6 $ 
\end{itemize}
with $\theta_{p}$ the polar angle of the outgoing proton, which has a modulus of the three-momentum $|\vec{p}_p|$.
T2K presents also a slightly less restrictive selection criterion for CC0$\pi1p$ requiring the detection of a muon and proton with the following conditions \cite{Abe:2018pwo}:
\begin{itemize} 
\item  $0.45~{\rm GeV} < |\vec{p}_p|<  1.0 {\rm ~GeV}$.
\item  $ \cos{\theta_{p}} > 0.4 $ 
\end{itemize}

\begin{figure}
\begin{center}
\includegraphics[width=0.495\textwidth]{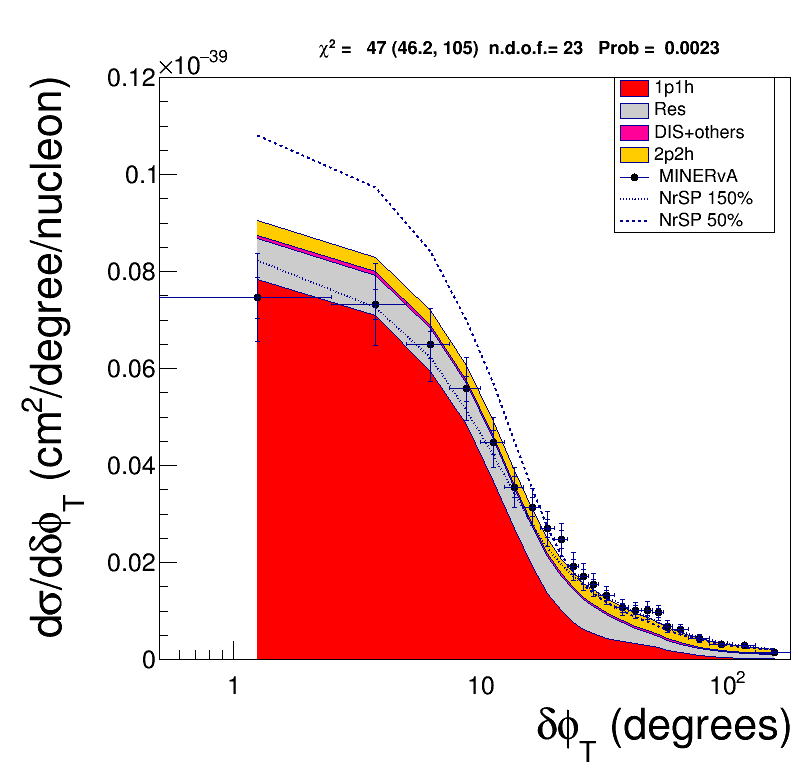}
\includegraphics[width=0.495\textwidth]{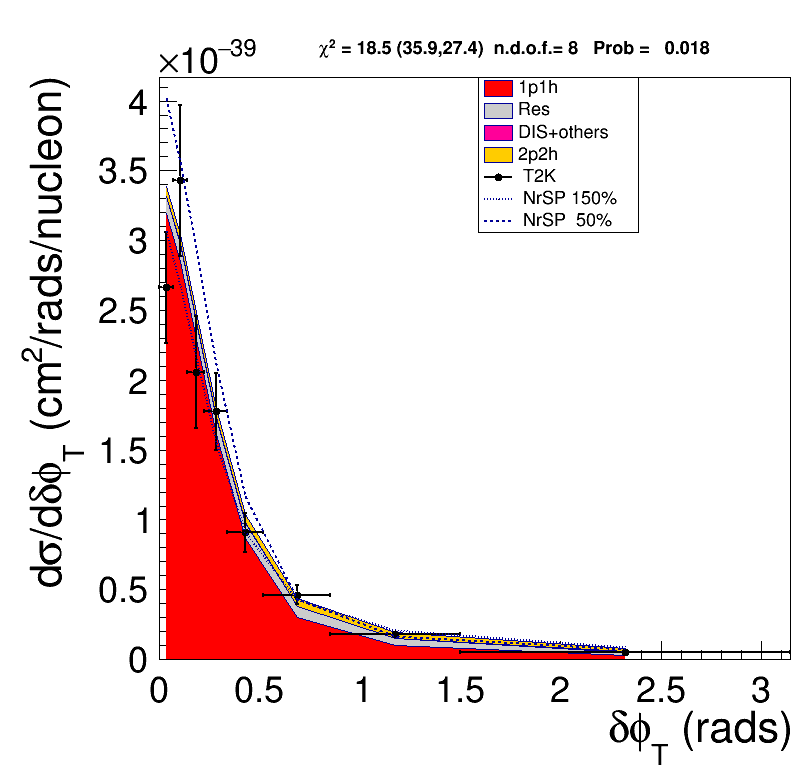}
\caption{\label{Fig:Dphit}   CC$0\pi 1p$ MINERvA~\cite{Lu:2018stk,Cai:2019hpx, Harewood:2019rzy} (left) and T2K~\cite{Abe:2018pwo} (right)  distributions for the TKI  angular variable $\delta \phi_T$. Details of the comparison with NEUT results as in Fig.~\ref{Fig:Dalphat}.   }
\end{center}
\end{figure}

\subsubsection{MINERvA data samples }

MINERvA, with an average neutrino energy of 3.5 GeV, published different event selections: the CC inclusive, the CC0$\pi$ that removes events with detected charged $\pi$'s and electromagnetic activity to eliminate  events with $\pi^0$, and the CC$0\pi1p$ sample which is a sub-sample of CC0$\pi$ requesting the presence of an identified proton in the final state. Another data sample from MINERvA, the so-called available energy \cite{Rodrigues:2015hik}, requires full simulation of the detector simulation that falls beyond the capabilities of this work. 

The CC inclusive selection in MINERvA requires \cite{Filkins:2020xol} a muon with polar angle ($\theta_{\mu}$) below $20^o$. The selection criteria for CC$0\pi$  demands  in addition~\cite{Ruterbories:2018gub}: 
\begin{itemize} 
\item a muon with momentum $|\vec{p}_{\mu}|$ between 1.5~GeV and 10~GeV,
\item neither charged, nor neutral pions escaping the nucleus. 
\item no $\gamma$ with energies about 10~MeV. We check this cut actually do not affect the MC prediction for CCQE and CC2p2h. 
\end{itemize}

The selection criteria for CC$0\pi1p$ requires in addition \cite{Lu:2018stk,Cai:2019hpx, Harewood:2019rzy}: 
\begin{itemize} 
\item a proton with polar angle ($\theta_{p}$)  below $70^o$.
\item a proton with momentum between 0.45~GeV and 1.2~GeV.
\end{itemize}

These two conditions are applied to protons leaving the nucleus after the  nucleon re-scattering.

\subsection{ Target composition } 

MINERvA and T2K targets are composed materials made of several components: CH, O, Al,.. Proportions in weight and nuclear content are given in Tables \ref{Tab:MinervaNuclei} and \ref{Tab:T2KNuclei}. To simulate the experimental composition we take the approximation of selecting only the main 3 components: C,H and O in proportions given in the tables. This actually has an appreciable effect on the selection mainly because of the large Fermi momentum and nuclear radius of the oxygen affecting the nucleon transport in the nucleus, accounting
for secondary collisions. Both experiments have the same target material (plastic scintillators) with very similar composition. The correction to the cross-section prediction introduced by be oxygen contribution is estimated by our models to be at the order of a percent.  The exception to this treatment is the CC0$\pi$ cross-section that is reported by T2K for a pure carbon target \cite{Abe:2020uub} and not in hydrocarbon as for the CC inclusive and MINERvA.

\begin{table}
\begin{center}
\begin{tabular}{ |c|c|c|c|c|c|c| } 
\hline
Component & CH & O & Al & Si & Cl & Ti  \\
\hline
Weight (\%) & 95.02 & 3.18 & 0.26 & 0.27 & 0.55 & 0.69 \\
Nuclei (\%) & 96.71 & 2.63 & 0.13 & 0.13 & 0.21 & 0.19 \\
\hline
Nucleons (\%) & 95.04 & 3.18 & 0.26 & 0.27 & 0.55 & 0.69 \\
\hline
\end{tabular}
\caption{\label{Tab:MinervaNuclei} MINERvA target nuclear composition in weight fraction and the translation to the fractional composition in nuclei.   Last row shows the fraction of nucleons of each of the spices with respect to the total.  Data taken from Ref.~\cite{Aliaga:2013uqz}}
\end{center}
\end{table}

\begin{table}
\begin{center}
\begin{tabular}{ |c|c|c|c|c| } 
\hline
Component & CH & O & Si & Ti  \\
\hline
Weight (\%) & 95.02 & 3.18 & 0.27 & 0.69 \\
Nuclei (\%) & 97.04 & 2.64 & 0.12 & 0.19  \\
\hline
Nucleons (\%) & 95.82 & 3.21 & 0.27 & 0.70 \\
\hline
\end{tabular}
\caption{\label{Tab:T2KNuclei} The same as Table~\ref{Tab:MinervaNuclei}, but for the T2K experiment. Data taken from Ref.~\cite{Amaudruz:2012esa}}
\end{center}
\end{table}

\begin{figure}
\begin{center}
\includegraphics[width=0.495\textwidth]{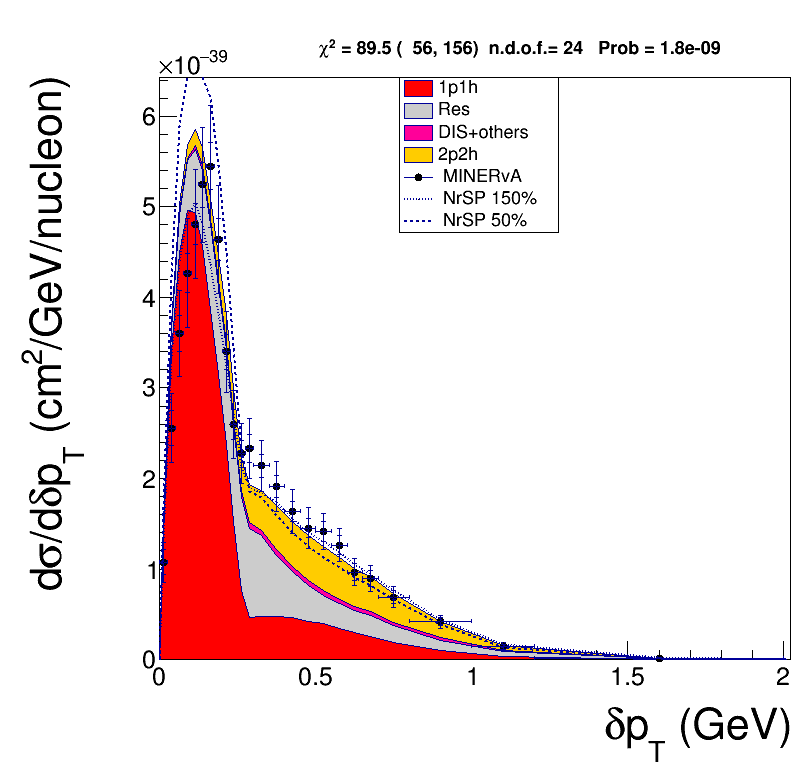}
\includegraphics[width=0.495\textwidth]{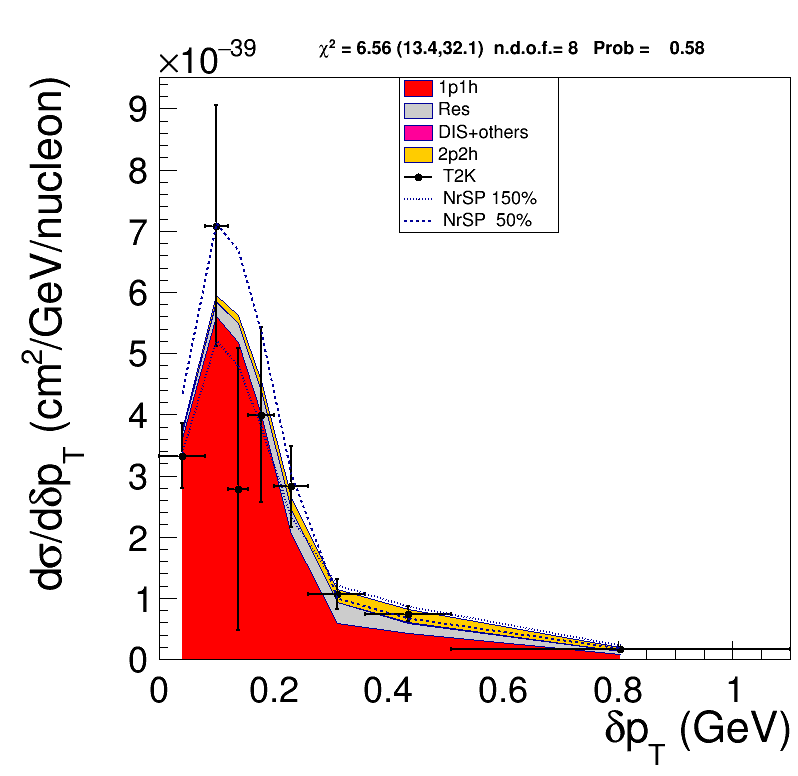}
\caption{\label{Fig:Dpt}  CC$0\pi 1p$ MINERvA~\cite{Lu:2018stk,Cai:2019hpx, Harewood:2019rzy} (left) and T2K~\cite{Abe:2018pwo} (right)  distributions for the missing transverse momentum  $|\delta\vec{p}_T|$. Details of the comparison with NEUT results as in Fig.~\ref{Fig:Dalphat}. }
\end{center}
\end{figure}

\begin{figure}
\begin{center}
\includegraphics[width=0.495\textwidth]{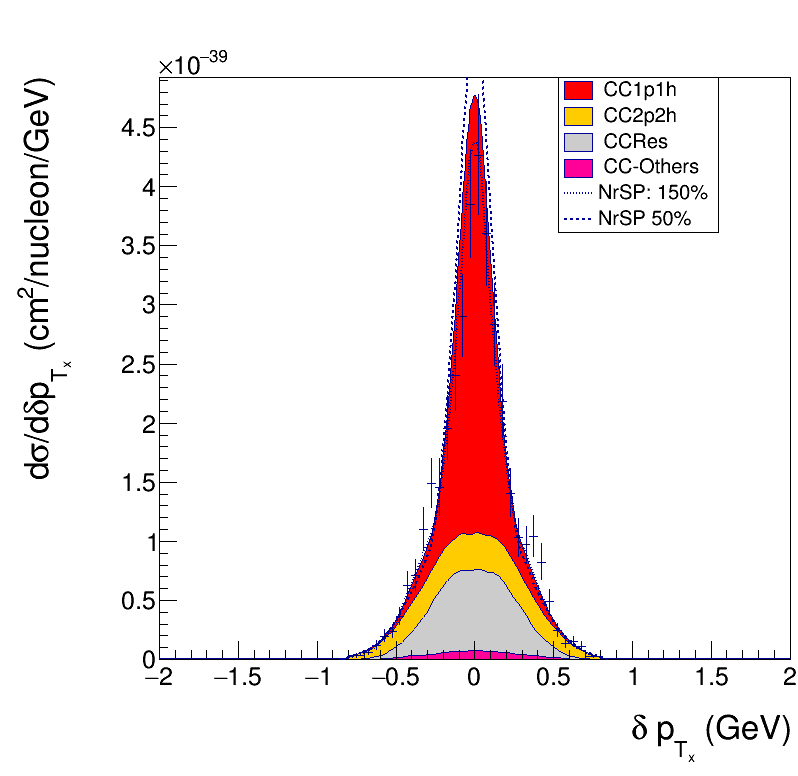}
\includegraphics[width=0.495\textwidth]{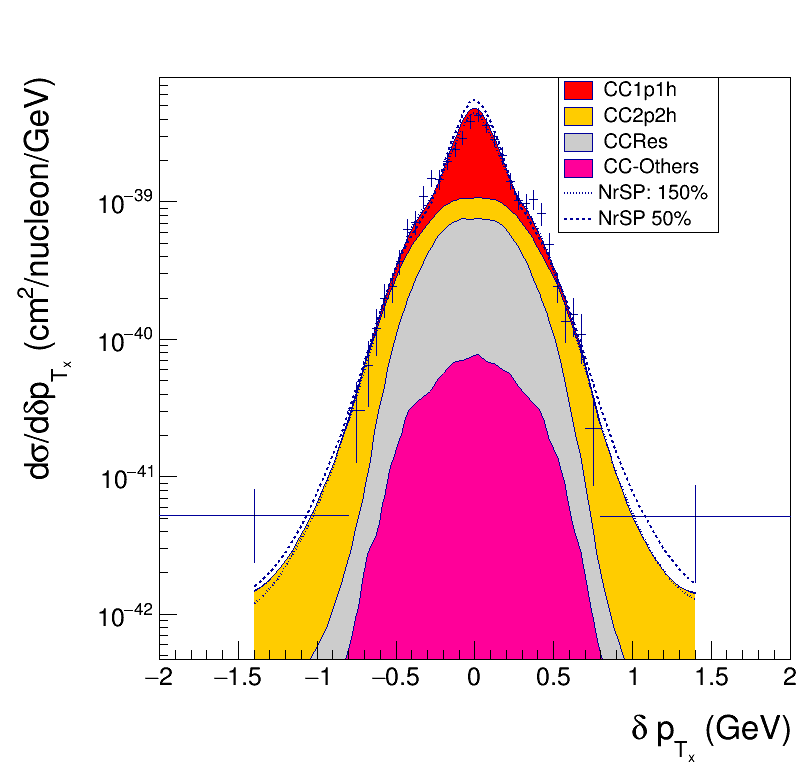}
\includegraphics[width=0.495\textwidth]{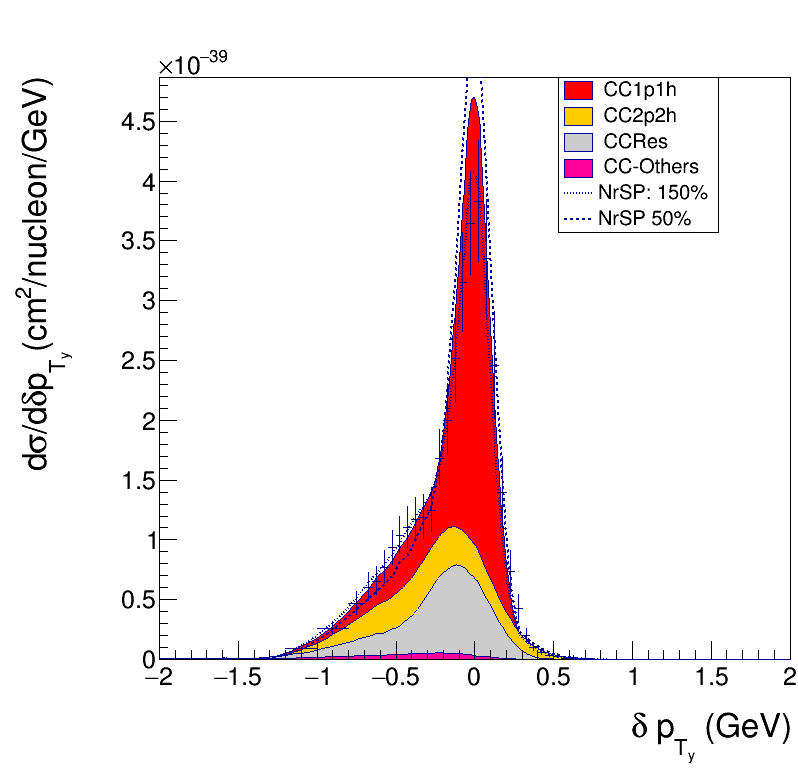}
\includegraphics[width=0.495\textwidth]{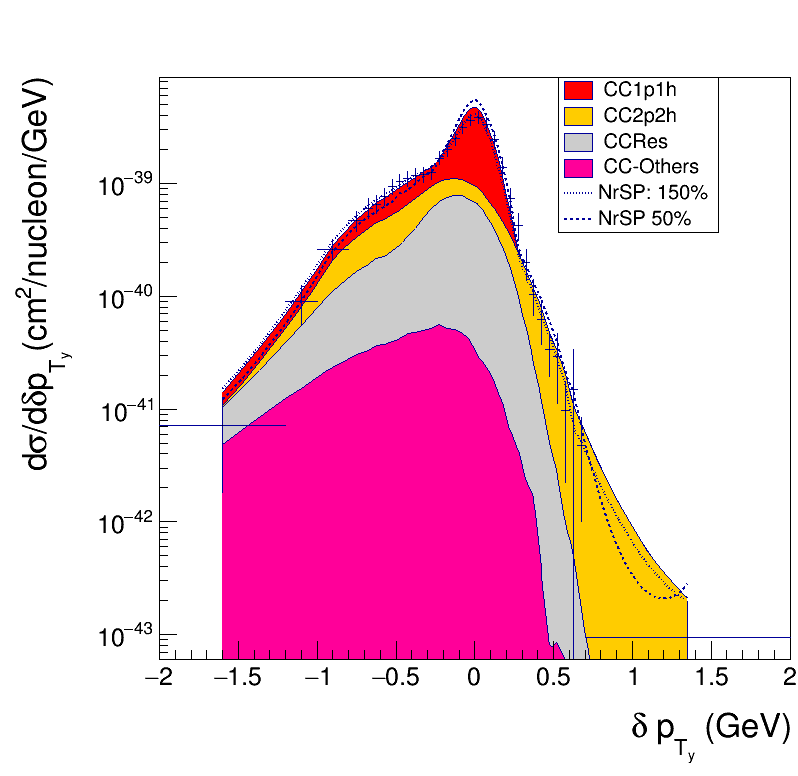}
\caption{\label{Fig:txy_Minerva} MINERvA $\delta p_{T_x}$ (top) and $\delta p_{T_y}$ (bottom) distributions~\cite{Lu:2018stk,Cai:2019hpx, Harewood:2019rzy}. Left and right  panels show the comparison (details as in Fig.~\ref{Fig:Dalphat}) with NEUT in linear and logarithmic scales, respectively.}
\end{center}
\end{figure}

\subsection{ Inclusive cross-sections } 
\label{sec:ICS}
In this section we compare our model predictions with the CC inclusive  and CC0$\pi$ measurements. The selection of events in the first reaction relies on the muon kinematics and ignore the hadronic component of the interaction. This sample allows to make a data-to-model comparison with a reduced bias from selection and detector acceptances, but it relies on the proper description of other interaction channels in the Monte Carlo. The CC0$\pi$ sample requires that there are no pions observed in the final state.  This selection, reduces the contribution of channels beyond CC1p1h and CC2p2h but it is affected by the correct modelling of both, the pion re-interactions inside the nucleus and  the primary pion production mechanism.

T2K published both the CC inclusive~\cite{Abe:2018uhf} and the CC0$\pi$~\cite{Abe:2020uub} cross-sections as a double differential distributions as function of the muon momentum and angle. The comparisons of these data-samples and the NEUT model predictions are shown in the  left and right panels of Fig.~\ref{Fig:CCincT2K}, respectively. The CC inclusive spectrum shows a sizeable contribution from resonant and deep inelastic scattering (DIS) channels, while for the CC0$\pi$ such components are considerably reduced, see also Table \ref{Tab:SampleComposition}. The remaining CCRes contribution comes from pion absorption in the nucleus. The model predicts reasonably well the tendencies in the experimental data. 
    
     MINERvA published the  muon 2D longitudinal and transverse momentum distributions for the CC inclusive cross-section measurements~\cite{Filkins:2020xol} (see top panels of Fig.~\ref{Fig:MinervaCCPT}).  The cross-section is dominated by DIS, but there are regions ($1.5~\rm{GeV}~ < p_{\parallel} < 5.0 $~GeV) where the QE and 2p2h contributions are relevant. The agreement with the present version of NEUT in the regions  where CCQE is relevant is  good, while the high longitudinal momentum histograms show cross-section predictions lower than data. The CC0$\pi$ sample from MINERvA is also shown in  Fig.~\ref{Fig:MinervaCCPT} (bottom panels). 
     It can be observed that the predictions are qualitatively similar to those found for the MINERvA CC inclusive case, lower than the data for large ($p_{\parallel} > 5.0$ GeV), but also for low ($p_{\parallel} < 3.5$ GeV) longitudinal momenta.

    To estimate the agreement data-model, we computed the $\chi^2-$merit function between the model predictions and data using the full error covariance matrices provided by the T2K and MINERvA experiments. The results are compiled in Tables~\ref{Tab:T2KChi2} and \ref{Tab:MinervaChi2}, respectively. The comparison is done for three different scale factors of the proton-nucleon cross-sections accounting for secondary collisions. Absolute $\chi^2-$values for the T2K CC-inclusive reaction are close to the dof and they show very little variations between the three examined scenarios. This is because re-scattering effects should not affect the inclusive cross section, and the observed differences should be produced by MC fluctuations (note that the three simulations are  statistically independent). The $\chi^2-$value found in this work is almost a factor of two smaller than those presented in the experimental paper (NEUT 5.3.2 and GENIE~\cite{Andreopoulos:2009rq} 2.8.0). The theoretical description of  the T2K CC$0\pi$ data sample  is better than that achieved for the inclusive one, with  $\chi^2-$value around 17 for 29 degrees of freedom. 
    In this case,  the different NrSP assumptions do not practically alter the results, as expected since the kinematics of the proton is not used in the CC$0\pi$ event selection. The $\chi^2$ value achieved with the present scheme is among the best reported in the T2K paper and it approaches the model "NEUT 5.4.1 LFG" in Table V of \cite{Abe:2020uub}, since both largely share the physics implementation.   
    
    On the other hand, the best-fit $\chi^2$ values for the CC inclusive MINERvA data sample, collected in Table \ref{Tab:MinervaChi2}, are among the best five reported in the experimental paper. Since this is dominated by the DIS cross-section, as discussed in Fig.~\ref{Fig:MinervaCCPT}, this result also confronts the prediction for this reaction channel in NEUT. The description of the MINERvA CC$0\pi$ events, where DIS has been practically removed, significantly improves, and it is quantitatively slightly poorer than that seen above for T2K. Nevertheless, the $\chi^2$ value is among the best two reported in the experimental paper and significantly better than any of the models including 2p2h contributions. NEUT predictions for MINERvA CC$0\pi$  show apparently a larger, though still soft, dependence on the proton re-scattering probability.

\begin{table}
\begin{center}
\begin{tabular}{ |c|c|c|c|c| } 
\hline
variable & dof & nominal & NrSP: 50\% & NrSP: 150\% \\
\hline
CC inclusive & 71 & 110 & 104 & 104 \\
CC$0\pi$ inclusive & 29 & 17 & 17 & 17 \\
\hline
$\delta \alpha_T $ & 8 & 35 & 29 & 43 \\
$\delta \phi_T $ & 8 & 19 & 27  & 36 \\
$\delta p_T $ & 8 & 7 & 32 & 13 \\
$\big |\Delta \vec{p}\,\big|$ & 49 & 261 & 193 & 398 \\
$\Delta \theta$ & 35 & 1021 & 798 & 1322 \\
$\Delta |\vec{p}\,|$ & 49 & 130 & 111 & 177 \\
\hline
\end{tabular}
\caption{\label{Tab:T2KChi2} $\chi^2-$likelihood test  for T2K variables. The number of degrees of freedom (dof) is given in the second column, while in the another three ones, results obtained for different distributions and three  NrSP --proton re-scattering probability-- configurations (nominal, 50\% and 150\% strengths) are collected.}
\end{center}
\end{table}
\begin{table}
\begin{center}
\begin{tabular}{ |c|c|c|c|c| } 
\hline
variable & dof & nominal & NrSP: 50\% & NrSP: 150\% \\
\hline
CC inclusive & 144 & 420 & 435 & 420 \\
CC0$\pi$  & 144 & 208 & 247 & 213 \\
\hline
$\delta \alpha_T $ & 12 & 21 & 25 & 18 \\
$\delta \phi_T $ & 23 & 47 & 105  & 46 \\
$\delta p_T $ & 24 & 89 & 155 & 56 \\
$\delta p_{T_x}$ & 32  & 63 & 106 & 43    \\
$\delta p_{T_y}$ & 33  & 56 & 108 & 40  \\
$ |\vec{p}_{\mu}| $ & 32 & 29 & 41 & 26\\
$ \theta_{\mu} $ & 19 & 23 & 24 &  22 \\
$ p_{p} $ & 25 & 30 & 33 & 31 \\
$ \theta_{p} $ & 26 & 49  & 62 & 43 \\
$ p_{n} $ & 24 & 107  & 202 & 86 \\
\hline
\end{tabular}
\caption{\label{Tab:MinervaChi2} Same as in Table~\ref{Tab:T2KChi2}, but for  MINERvA experiment. }
\end{center}
\end{table}

\subsection{ Transverse kinematic-imbalance (TKI) variables }
\label{sec:TV}

\begin{figure}
\begin{center}
\includegraphics[width=0.495\textwidth]{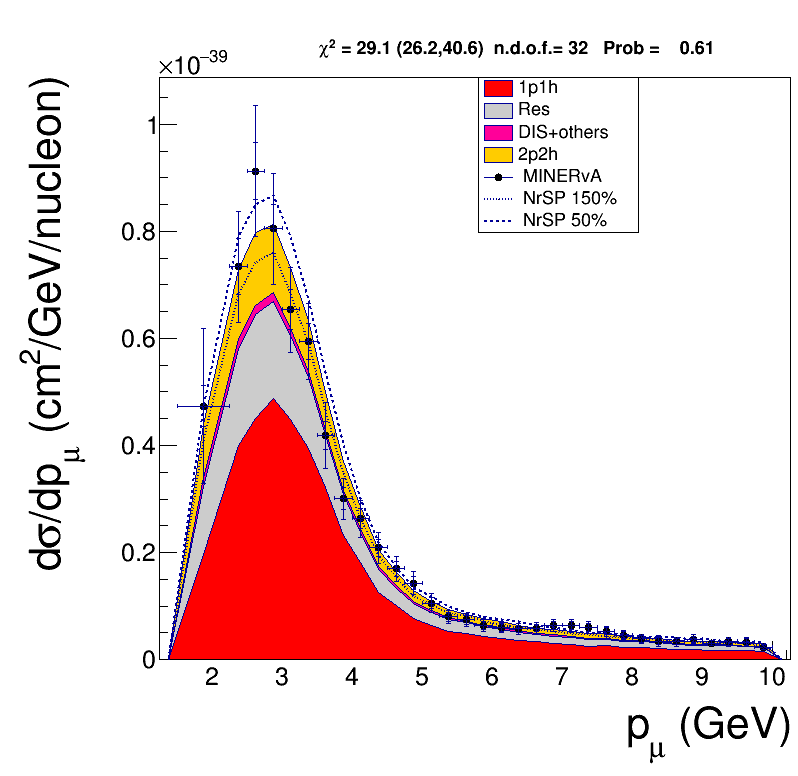}
\includegraphics[width=0.495\textwidth]{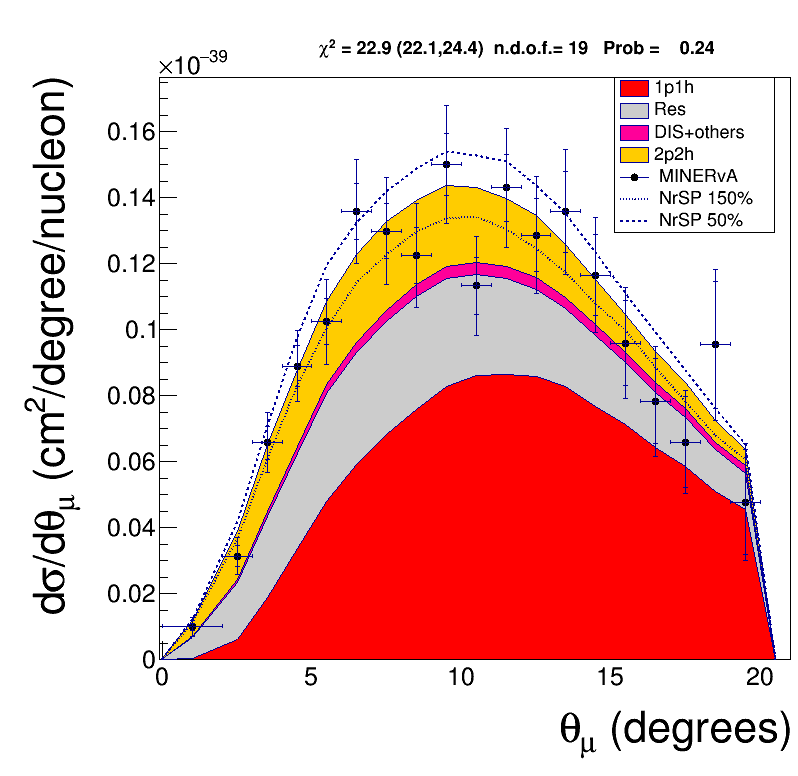}
\caption{\label{Fig:Pmu_Minerva} Inclusive $|\vec{p}_{\mu}|$ (left) and $\theta_{\mu}$ (right) distributions from the CC$0\pi 1p$ MINERvA sample~\cite{Lu:2018stk,Cai:2019hpx, Harewood:2019rzy}. Details of the comparison with NEUT results as in Fig.~\ref{Fig:Dalphat}. }
\end{center}
\end{figure}
\begin{figure}
\begin{center}
\includegraphics[width=0.495\textwidth]{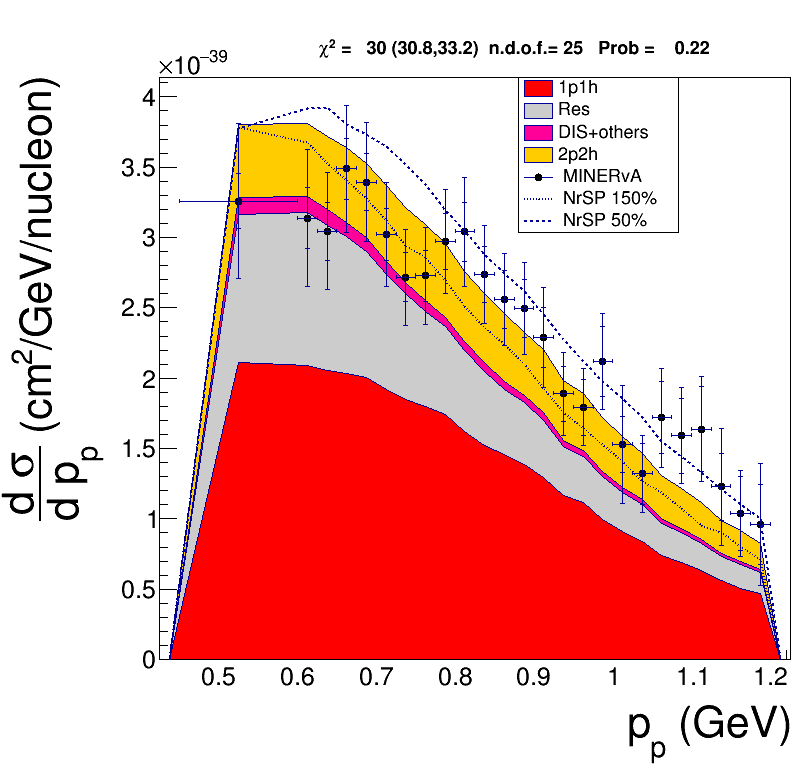}
\includegraphics[width=0.495\textwidth]{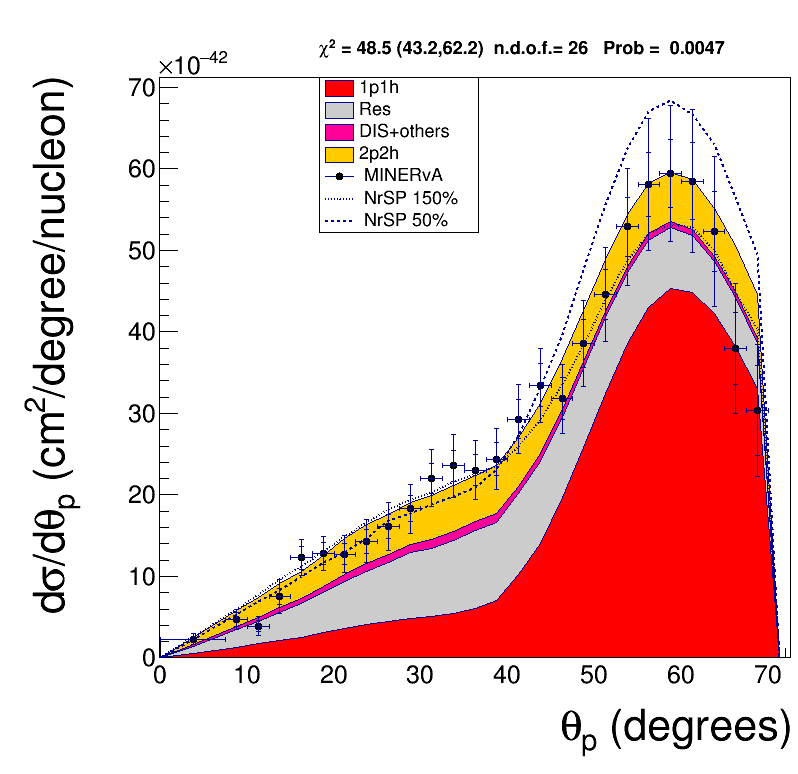}
\caption{\label{Fig:Pp_Minerva} MINERVA CC0$\pi$1p differential cross sections~\cite{Lu:2018stk, Cai:2019hpx, Harewood:2019rzy} in proton kinematics, momentum ($|\vec{p}_p|$) and angle ($\theta_{p}$),  together with results from the current implementation of NEUT. Details of the comparison with NEUT as in Fig.~\ref{Fig:Dalphat}.}
\end{center}
\end{figure}

\begin{figure}
\begin{center}
\includegraphics[width=0.5\textwidth]{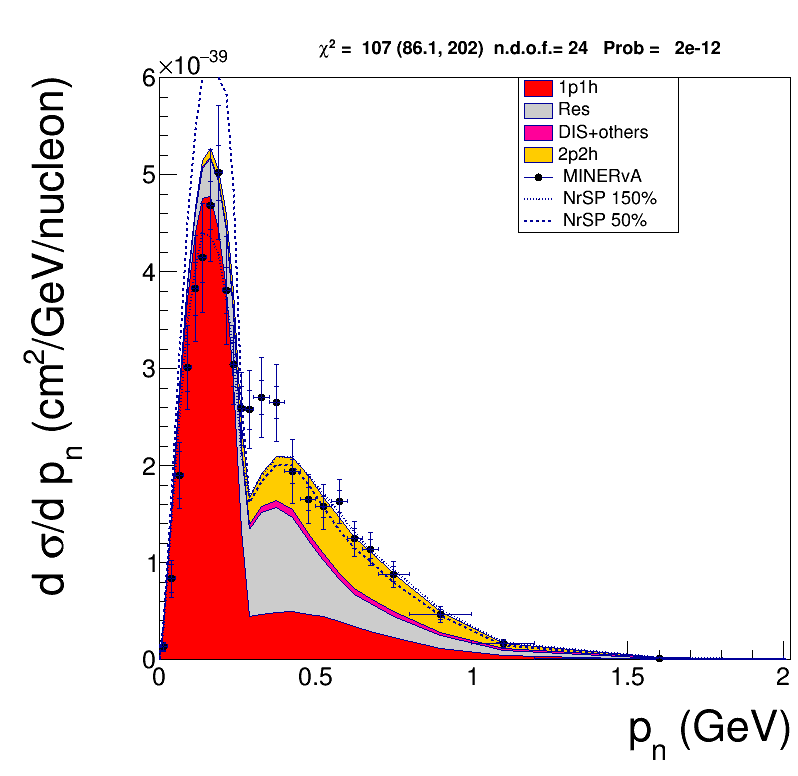}
\caption{\label{Fig:Pn_Minerva} MINERvA CC0$\pi$1p differential cross sections in reconstructed $|\vec{p}_n|$. Data from Ref.~\cite{Lu:2018stk, Cai:2019hpx, Harewood:2019rzy}. We also show  results from the current implementation of NEUT, with details of the comparison as in Fig.~\ref{Fig:Dalphat}.}
\end{center}
\end{figure}

TKI variables~\cite{Lu:2015hea, Lu:2015tcr}, depicted in Fig.~\ref{Fig:transverse}, lead to observable distributions with minimal
dependence on neutrino energy, which provide direct constraints on nuclear effects in (anti)neutrino-nucleus interactions. Obviously, the measurement of the distribution of these variables require the observation of the outgoing proton (neutrino reactions) or the neutron (anti-neutrino reactions) in the final state. The missing transverse, with respect to the neutrino direction, momentum is computed as
\begin{equation}
 |\delta\vec{p}_T| \equiv |\vec{p}_{T}|= \frac{ \left|(\vec{p}_{\mu}+\vec{p}_{p})  \times \vec {p}_{\nu} \right|}{| \vec{p}_{\nu} | }   
 \end{equation}
In addition, the missing transverse momentum is separated in two components in relation to the $\nu-\mu$ reaction plane,  defined by the  neutrino and the emitted charged lepton. One of the transverse components is contained in the reaction plane $\delta p_{T_y}$, while the another one, $\delta p_{T_x}$, is perpendicular to this plane~\cite{Cai:2019hpx}.  

On the other hand, the TKI angular variables read: 
\begin{eqnarray}       
 \delta \phi_T &=& \arccos\left[ \frac{-\vec{p}_{T_{p}} \cdot \vec{p}_{T_{\mu}}}{ | \vec{p}_{T_{p}} |  |\vec{p}_{T_{\mu}}|} \right] \\
 \delta \alpha_T &=&  \arccos\left[ \frac{- \vec{p}_{T} \cdot \vec{p}_{T_{\mu}}}{| \vec{p}_{T} | \cdot |\vec{p}_{T_{\mu}}|} \right] 
\end{eqnarray}
 where $\vec{p}_{T}=\vec{p}_{T_{\mu}}+\vec{p}_{T_{p}}$, with $\vec p_{T_{\mu}}$ (or $\vec{p}_\perp$, as used in Fig.~\ref{Fig:MinervaCCPT}) and $\vec{p}_{T_{p}}$  the transverse projections of the muon and proton momenta to the neutrino direction. This discussion focuses on the QE-like process $\nu_\mu + A \to \mu^- + p +X$, where $X$ is a final-state hadronic system consisting of the nuclear remnant with possible additional protons but without pions that indicate resonant or other processes. There is an imbalance, $\delta\vec{p}$, between the initial neutrino momentum and the sum of final-state lepton and hadron momenta as a result of nuclear effects.
Under the assumption that $X$ is just the remnant nucleus, with $(A-1)$ nucleons,  then $|\delta\vec{p}\,|$ gives the magnitude of its recoil
momentum, which can be obtained~\cite{Furmanski:2016wqo} independently of the  unknown incident neutrino energy. Moreover, assuming perfect balance of momentum in the reaction (see discussion of Eqs.~\eqref{eq:mom_con} and \eqref{eq:balmon}), $|\delta\vec{p}\,|$ can be identified to the neutron target momentum, which is then given in terms of  measurable quantities~\cite{Lu:2018stk,Furmanski:2016wqo}
\begin{equation}
     |\vec{p}_n| = \sqrt{  |\vec{p}_{T}|^2  +  |\vec{p}_{L}|^2   } 
\end{equation}
with  
 \begin{equation}   
 |\vec{p}_{L}| = \frac{ (M_A - E_{\mu} - E_{p}^\infty + |\vec{p}_{L_p}|)^2  -(M'_{A-1})^2 - |\vec p_{T}|^2 }{ 2 (M_A - E_{\mu} - E_{p}^\infty +  |\vec{p}_{L_p}| + |\vec{p}_{L_\mu}|)} 
 \end{equation}
where  $\vec{p}_{L_p}$ and $\vec{p}_{L_\mu}$ (or $\vec{p}_\parallel$, as used in Fig.~\ref{Fig:MinervaCCPT}) denote the projections of the corresponding three-vectors
on the direction of the incoming neutrino. For the MINERvA measurement of Ref.~\cite{Lu:2018stk, Cai:2019hpx, Harewood:2019rzy}, $M_A$ is the mass of the carbon target, while  for $M'_{A-1}$ is taken  $(M_A - m_{n} + E_b)$, with $E_b$ the nucleon binding energy that is fixed to 27.13~MeV and $m_{n}$ is the mass of the neutron\footnote{Note, this is the prescription used in the experimental work, which should be used to compare to the event-distributions provided in that paper, but it does not correspond to the energy-balance of Eq.~\eqref{eq:ene-bal} proposed here for genuine QE processes.}.

   The T2K collaboration considered  additional distributions based on the neutrino energy reconstruction formula used in neutrino oscillation experiments. This prescription assumes a genuine QE event where the target nucleon is at rest, and the neutrino energy is reconstructed assuming energy and momentum conservation (see for instance Ref.~\cite{Nieves:2012yz}): 
   \begin{equation} 
   E_{\nu}^{\rm rec} = \frac{ m_p^2-m_{\mu}^2+2 E_{\mu}(m_n-E_b)-(m_n-E_b)^2}{ 2 ((m_n-E_b)-E_{\mu}+|\vec{p}_{\mu}|\cos{\theta_{\mu}}) }
   \end{equation}
   where $\theta_{\mu}$ is the angle of the muon with respect to the average neutrino direction, $m_{\mu}$ is the mass of the muon,  $m_p$ is the mass of the outgoing proton  and the constant $E_b$ is fixed to 25~MeV in the experimental T2K results~\cite{Abe:2018pwo}. Once the neutrino energy is known, one can compute the so-called inferred proton momentum: 
   \begin{equation} 
   \vec{p}_{p}^{\,\,\rm inf} = \vec{p}^{\,\,\rm rec}_{\nu} - \vec{p}_{\mu}
   \end{equation} 
   New observables are build by comparing the inferred momentum with the experimentally measured proton momentum ($\vec{p}_p^{\, \infty}$): 
   \begin{eqnarray}
   | \Delta \vec p\, |  &=& | \vec p_{p}^{\,\,\rm inf} - \vec p_{p}^{\,\infty}| \\
   \Delta |\vec p\, |  &=&  |\vec p_{p}^{\,\,\rm inf}| - |\vec p_{p}^{\, \infty}| \\
   \Delta \theta  &=&  \arccos\left[ \frac{\vec p_{p}^{\, \infty} \cdot \vec p_{\nu} }{|\vec p_{p}^{\, \infty} ||\vec p_{\nu} |}\right] - \arccos\left[ \frac{\vec p_{p}^{\,\, \rm inf} \cdot \vec p_{\nu}}{|\vec p_{p}^{\,\, \rm inf}| |\vec p_{\nu} |}\right] 
    \end{eqnarray}  
 
    The MINERvA Collaboration reported different cross-sections from its CC$0\pi 1p$ data sample~\cite{Lu:2018stk, Cai:2019hpx, Harewood:2019rzy} : 
    
    \begin{itemize}
        \item Inclusive lepton momentum and angle. 
        \item Reconstructed kinematics such  as  the target nucleon momentum $|\vec{p}_n|$.
        \item Visible proton momentum. 
        \item TKI angles: $\delta \phi_T$ and $\delta \alpha_T$.
        \item Transverse momentum balance: $\delta  p_T$, $\delta p_{T_x}$, $\delta p_{T_y}$.
    \end{itemize}
    
    while  T2K reported on slightly different set of variables from its CC$0\pi 1p$ sample~\cite{Abe:2018pwo}: 
    
       \begin{itemize}
        \item Transverse angles (TKI variables with first selection criterion): $\delta \phi_T$ and $\delta \alpha_T$, and  the missing transverse momentum  $|\delta\vec{p}_T|$ 
        \item Proton momentum balance (Inferred variables with the second selection criterion): $|\Delta \vec p\,|$,  $\Delta \theta$ and $\Delta |\vec p\,|$ in bins of muon momentum and angle. 
    \end{itemize}

Experimental and  NEUT predictions for the TKI angular variables $\delta \alpha_T$ and $\delta \phi_T$, and  the missing transverse momentum  $|\delta\vec{p}_T|$, both for MINERvA (left) and T2K (right) are shown in Figs.~\ref{Fig:Dalphat}, \ref{Fig:Dphit} and \ref{Fig:Dpt}, respectively. Absolute $\chi^2-$values obtained from the comparison with data are compiled in Tables~\ref{Tab:T2KChi2} and \ref{Tab:MinervaChi2}, respectively.  The NEUT results for the TKI variables are in general in an acceptable agreement with both MINERvA and T2K data, though the MINERvA distributions are better described. As expected, we observe some significant dependence on the proton re-scattering probability, which is reflected in  the $\chi^2-$likelihood tests. We should point out that NrSP effects change not only the overall normalization, but also the shape of the distributions. This is clearly visible, for instance, in the CC$0\pi 1p$ T2K~\cite{Abe:2018pwo} $\delta \alpha_T$ distributions depicted in Fig.~\ref{Fig:Dalphat}. There, we see that re-scattering effects become more important in the region of smallest  $\delta \alpha_T$ angles. This observation is in order for some  other distributions discussed below in this subsection.
 
The transverse momentum components contain different information. The component  $\delta p_{T_x}$ is expected to be symmetric around zero, with a width which  depends on the target neutron momentum and  further re-scattering effects, while $ \delta p_{T_y}$ can be asymmetric due to leading effect of the neutrino boost. Results are shown in Fig.~\ref{Fig:txy_Minerva}. The tendency is very well described by the model with the long tails dominated by 2p2h, resonant and, DIS mechanisms in the $\delta p_{T_x}$ ($\delta p_{T_y}$) distribution. The $\chi^2$ comparison, see Table \ref{Tab:MinervaChi2} shows an excellent agreement with a preference for an increase in the NrSP, which reduces the contribution of the CCQE channel by reducing the probability of proton tagging in detectors. This tendency is shared by most of the other TKI observables. 

MINERvA collaboration also reported  $|\vec{p}_{\mu}|$  and $\theta_{\mu}$  distributions from its CC$0\pi 1p$ MINERvA sample~\cite{Lu:2018stk, Cai:2019hpx, Harewood:2019rzy}. The comparison of these data with the current implementation of NEUT is shown in  Fig.~\ref{Fig:Pmu_Minerva}. We find a quite good description of these two event distributions, with $\chi^2/dof$ around one (see Table~\ref{Tab:MinervaChi2}) for nominal NrSP, and some dependence on this latter input as expected when analysing CC$0\pi 1p$ data-samples. 

The experimental results with a visible proton in the final state are biased towards high momentum transfers since the proton should have at least 450~MeV to be detected. The typical proton momentum and angle distributions in MINERvA are shown in Fig.~\ref{Fig:Pp_Minerva}. On the contrary, the samples with additional invisible protons do not suffer from large momentum transfer biases. The difference between the two are dominated by low $|\vec{q}\,|$ contributions with feed-down background cause by NrSP. These tendencies can be observed in Figs.~\ref{Fig:Q0Q2MinervaT2K} and \ref{Fig:Q0Q2MinervaT2KCC0pi}  of Appendix~\ref{App:Momentum transfer}, where the model estimation for the energy ($q^0 = E_{\nu} - E_{\mu}$) and momentum  ($|\vec{q}\,| = | \vec p_{\nu} - \vec p_{\mu} |$) transfer distributions for the MINERvA and T2K CC inclusive, CC$0\pi$ and CC$0\pi$1p event selections are shown.

The overall agreement with NEUT is good, showing the importance of 2p2h mechanisms.  The results for the TKI variables are good for the MINERvA data, with statistically acceptable values of $\chi^2/dof$ for most of the cases. The worst comparison is obtained for the reconstructed $|\vec{p}_n|$ variable, where a large discrepancy is observed in the region around 0.3~GeV (see Fig.~\ref{Fig:Pn_Minerva}). This is at the transition from the CC1p1h dominated cross-section to the one dominated by resonance and CC2p2h mechanisms. This is actually the most distinctive difference in all the comparisons of this work and a nice reference observable to try model variations. As it is is shown in Fig.~\ref{Fig:Pn_Minerva} and Table~\ref{Tab:MinervaChi2},  the variation of the proton re-scattering probability does no alleviate the discrepancy. Since, the high momentum ($|\vec{p}_n| \ge 0.5$ GeV) region is well reproduced by the model, the discrepancy seems to be led by the transition, either from non described tails in the CC1p1h, which might come from high energy neutron target components predicted by realistic SFs, or by a miss-representation of resonant or CC2p2h models. In any case, it seems that a re-weight of the cross-section will not improve the agreement. 

\begin{figure}
\begin{center}
\includegraphics[width=\textwidth]{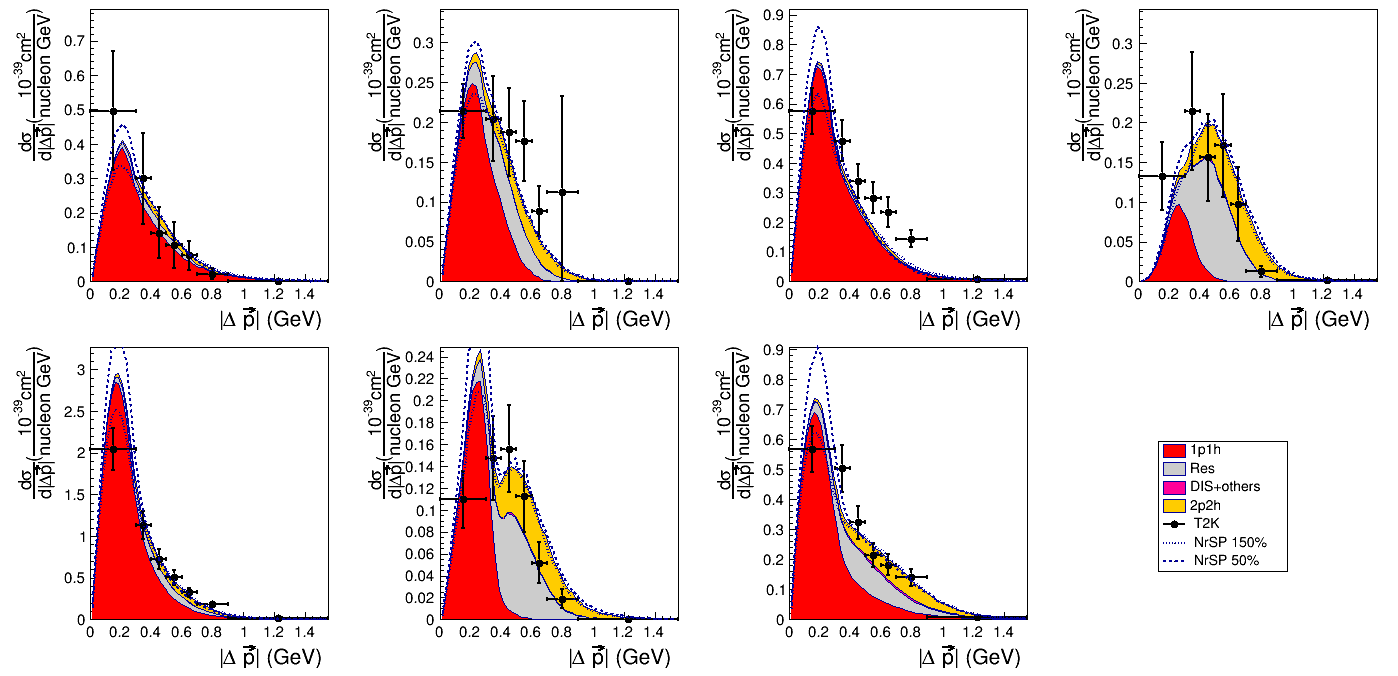}
\caption{\label{Fig:dp_T2K} T2K CC0$\pi$1p $|\Delta\vec{p}\,|$ distribution. The panels correspond to different muon kinematic bins. From left to right and up to down: $-1<\cos\theta_\mu < -0.6$, $-0.6<\cos\theta_\mu < 0$ with $|\vec{p}_\mu|< 250$ MeV, $-0.6<\cos\theta_\mu < 0$ with $|\vec{p}_\mu|> 250$ MeV,  $0<\cos\theta_\mu < 1$ with $|\vec{p}_\mu|< 250$ MeV, $0<\cos\theta_\mu < 0.8$ with $|\vec{p}_\mu|> 250$ MeV, $0.8<\cos\theta_\mu <  1$ with $250 ~{\rm MeV} < |\vec{p}_\mu|< 750$ MeV and $0.8<\cos\theta_\mu < 1$ with $|\vec{p}_\mu|> 750$ MeV. Data taken from Ref.~\cite{Abe:2018pwo}. \label{fig:T2K-CC0pi1p-1}}
\end{center}
\end{figure}
\begin{figure}
\begin{center}
\includegraphics[width=\textwidth]{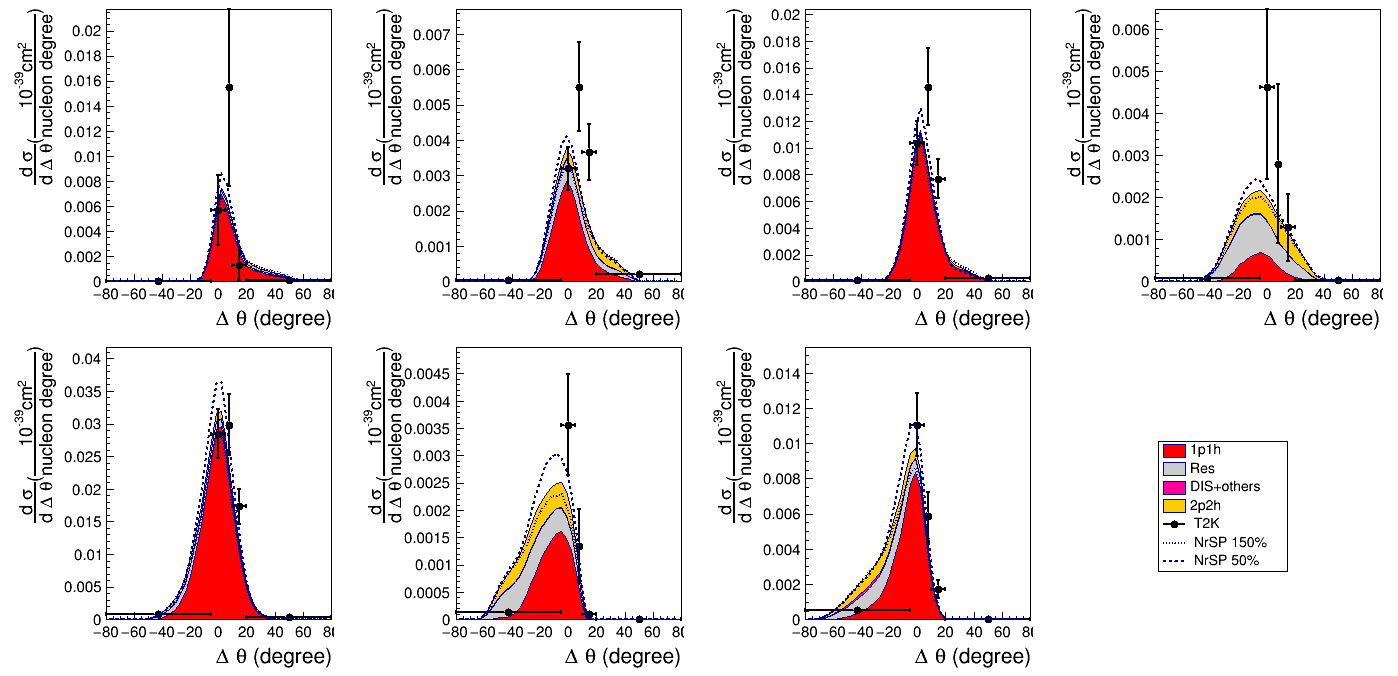}
\caption{\label{Fig:dtheta_T2K} T2K  CC0$\pi$1p $|\Delta\theta|$ distribution. The panels correspond to the  muon kinematic bins specified in Fig.~\ref{Fig:dp_T2K}. Data taken from \cite{Abe:2018pwo}. \label{fig:T2K-CC0pi1p-2} }
\end{center}
\end{figure}
\begin{figure}
\begin{center}
\includegraphics[width=\textwidth]{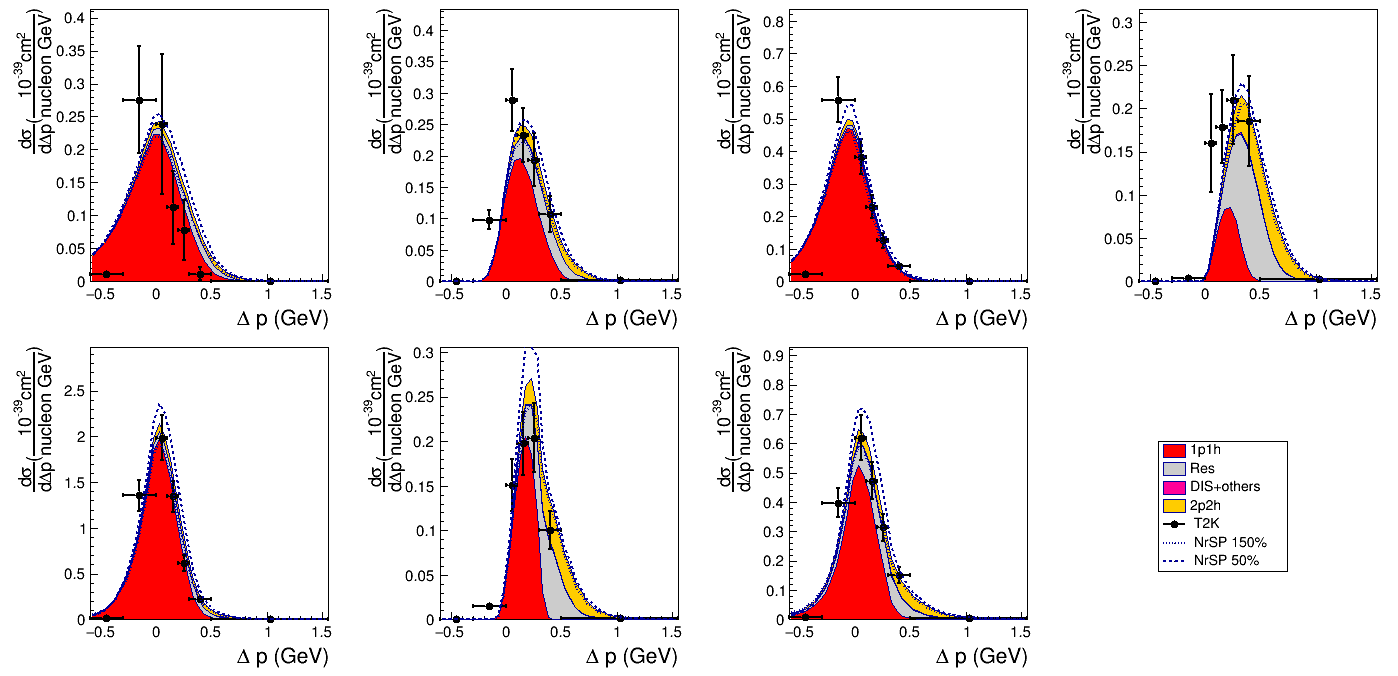}
\caption{\label{Fig:dabsp_T2K} T2K CC0$\pi$1p $\Delta |\vec{p}\,|$ distribution.  The panels correspond to the  muon kinematic bins specified in Fig.~\ref{Fig:dp_T2K}. Data taken from \cite{Abe:2018pwo}.\label{fig:T2K-CC0pi1p-3} }
\end{center}
\end{figure}

The agreement with T2K CC0$\pi$1p data is less impressive and the obtained $\chi^2/dof$, see Table \ref{Tab:T2KChi2}, are large for all the three observables $|\Delta\vec{p}\,|$, $|\Delta\theta|$ and $\Delta |p|$ reported in \cite{Abe:2018pwo}, and shown here in Figs.~\ref{fig:T2K-CC0pi1p-1}--\ref{fig:T2K-CC0pi1p-3}. The worst situation is found for the $\Delta\theta$ distribution, with the largest contributions to $\chi^2$ produced by the negative $\Delta\theta$ bins. There,  the number of events is always very small independent of the muon-kinematics, and the present model fails to properly describe those data, though  one should bear in mind that these bins have a negligible weight in the totally integrated  cross section. Indeed, if these bins are removed,  the merit-function $\chi^2$ is reduced to 66 for 30 degrees of freedom.  The agreement is slightly better for the re-scaling factor NrSP=0.5 that reduces the scattering of the outgoing proton in the nucleus (54 for negative values of $\Delta\theta$). The $\chi^2$ figures for $|\Delta\vec{p}\,|$, and $\Delta |p|$ are also reduced with the re-scaling factor NrSP=0.5.

 In summary from the results of Tables \ref{Tab:T2KChi2} and \ref{Tab:MinervaChi2}, we conclude that MINERvA TKI distributions are better described with the 150\% enhanced NrSP, while the T2K ones favor either nominal ($\delta \phi_T$ and $\delta p_T $) or the 50\% reduced  NrSP ($\delta \alpha_T $, $\big |\Delta \vec{p}\,\big|$, $\Delta \theta$ and  $\Delta |\vec{p}\,|$) configurations.

 The difference between the agreement found for MINERvA and T2K data samples  might point to an energy dependent deviation. 
The LFG model could provide a better approximation to the MINERvA energies than to the T2K ones, which would be more sensitive to finer details of the low-lying nuclear levels.  However, an overall $\chi^2-$analysis is not sufficient to extract robust conclusions on the energy dependence of the LFG model, and it should also be noted that, as discussed above,  MINERvA and T2K data-sets favor different proton re-scattering configurations. This could be an indication that the different agreement exhibited for the MINERvA and T2K data-samples might be also partially  produced by higher energy channels such as 2p2h or resonance mechanisms. At this respect, it would be very useful to have access to 
the T2K $d \sigma/ dp_{n}$ cross-section, since this distribution is specially sensitive to non CC1p1h contributions, as seen in Fig.~\ref{Fig:Pn_Minerva} for MINERvA data.
 
\subsection{ Integrated cross-section } 
 
 We have performed the numerical integrals of the differential distributions for the six event samples examined in this work. The obtained cross sections, both from data and from the model predictions are compiled in Table~\ref{Tab:IntegratedCrossSection}. The results show a common tendency of the NEUT model to predict lower cross-sections for the inclusive and the CC0$\pi$ samples, while its predictions are marginally larger  for the CC0$\pi$1p data-set. The deviations with the theoretical approach for both experiments are similar, except for the case of the CC inclusive. This is expected due to the DIS cross-section, and the very different proportion predicted for this channel for the MINERvA and T2K experiments, see Table~\ref{Tab:SampleComposition}.  The change in the Exp to Model ratio observed from the CC0$\pi$ and the CC0$\pi$1p may be a consequence of both the proton momentum prediction below the detector detection thresholds and the proton NrSP. The effect of the NRsP on the integrated CC0$\pi$1p cross-section is shown in Table~\ref{Tab:IntegratedCrossSectionNRsP}, and it is also included in the proportions collected in Table~\ref{Tab:SampleComposition}. The number of events with visible protons increases when reducing the NrSP and vice-versa. The effects on MINERvA are slightly reduced with respect to those found for T2K due to the larger proton momentum expected at higher neutrino energies. Even if a reduced NrSP choice will bring the numerical values closer to the differences seen for the CC0$\pi$ samples, we should be cautious. The effect of the CCRes model is not apparent in the results, but however, 23\%   in MINERvA and 12.5\% in T2K of the CC0$\pi$1p events come from CCRes according to our model. The modelling of the CCRes should also take into account the absorption of the emerging pions by the nucleus. 
 
 If, on the contrary, we assume that the CCRes is well simulated, the results point to a deficit in the model prediction of low momentum protons ($\le$ 450 MeV). A larger re-scattering probability would approach the two results. Another possible cause of the discrepancy is the larger $|\vec{q}\,|$ values intrinsic to CC0$\pi$1p with respect to CC0$\pi$ since the proton should be emitted with momentum greater that 450~MeV. 
 
 Nevertheless, given the experimental uncertainties also included in  Table~\ref{Tab:IntegratedCrossSection}, we conclude that the present model implemented in NEUT leads to reasonable integrated cross sections for all six  samples considered in this work.

\begin{table}
\begin{center}
\begin{tabular}{ |l|c|c|c| } 
\hline
&  Exp [$10^{-39}$ cm$^2$] & Model [$10^{-39}$ cm$^2$] & $ \frac{Exp-Model}{Model} $ [$\%$]\\
\hline
\multicolumn{4}{|c|}{MINERvA}  \\
\hline
CC inclusive & $18.3 \pm 1.3$  & $16.7$ & $ 9.8 \pm 7.8$ \\ 
CC0$\pi$ & $4.64 \pm 0.38$  &  $3.96$ & $17.1 \pm 9.7$ \\
CC0$\pi$1p from $\delta\alpha_T$  & $1.85 \pm 0.17$ & $1.94$ & $-4.6 \pm 9.0$ \\
\hline
\multicolumn{4}{|c|}{T2K} \\
\hline 
CC inclusive &  $7.33 \pm 0.81$  &  $6.01$ & $22 \pm 13$ \\
CC0$\pi$ & $4.74 \pm 0.59$  &  $3.78$ & $25 \pm 16$ \\
CC0$\pi$1p from $\delta\alpha_T$ & $1.39 \pm 0.13$  & $1.46$ & $-4.6 \pm 9.0$ \\
CC0$\pi$1p from $|\Delta \vec p\,|$ & $3.01 \pm 0.34$  & $2.88$ & $5 \pm 12$ \\
\hline
\end{tabular}
\caption{\label{Tab:IntegratedCrossSection} Integrated cross-sections for the different event selections provided by T2K and MINERvA. The CC0$\pi$1p integrated cross-section has been computed using the $d \sigma/ d (\delta\alpha_T)$ differential distribution for MINERvA and T2K and  also $d \sigma / d|\Delta \vec p\,|$ for T2K. In all cases, the nominal NrSP configuration is used.}
\end{center}
\end{table}
\begin{table}
\begin{center}
\begin{tabular}{ |l|c|c|c|c| } 
\hline
          & 1p1h  (\%) & 2p2h (\%) & Res  (\%) & DIS-others  (\%) \\
\hline     
\multicolumn{5}{|c|}{MINERvA}  \\
\hline
CC inclusive & 19.0 & 4.7 & 35.4 & 40.9 \\ 
CC0$\pi$ & 64.8 & 15.2 & 17.8 & 2.1 \\
CC0$\pi$1p  from $\delta \alpha_T$ & $56.8^{-1.4}_{+1.3}$ & $16.9^{+0.2}_{-0.3}$ & $23.5^{+1.1}_{-1.0}$ & $2.7^{+0.2}_{-0.1}$  \\
\hline
\multicolumn{5}{|c|}{T2K}  \\
\hline
CC inclusive & 46.9 & 6.2 & 31.8 & 15.0 \\ 
CC0$\pi$ & 78.8 & 10.4 & 10.0 & 0.8 \\
CC0$\pi$1p from $\delta \alpha_T$ & $80.2^{-1.0}_{+0.8}$ & $8.4^{+0.2}_{-0.1}$ & $10.8^{+0.7}_{-0.6}$ & $0.6^{+0.0}_{-0.0}$\\
CC0$\pi$1p from $|\Delta \vec p\,|$ & $77.0^{-1.2}_{+0.9}$ & $9.8^{+0.2}_{-0.1}$ & $12.7^{+0.9}_{-0.8}$ & $0.7^{+0.1}_{- 0.0}$ \\

\hline
\end{tabular}
\caption{\label{Tab:SampleComposition} Different NEUT contributions to the  T2K and MINERvA cross sections given in Table~\ref{Tab:IntegratedCrossSection}. As in this latter table, the two T2K CC0$\pi$1p selections have been considered
and all results are obtained using the nominal NrSP. For those samples with explicit proton in the final state (CC0$\pi$1p), we compute the variation of the proportions induced by the use of stronger or weaker proton re-scattering. The differences with respect to the results obtained from the nominal NrSP configuration are shown as an upper  (NRsP 150\%) and a lower (NRsP 50\%) error. As expected, the increase (decrease) in the NRsP strength reduces (increases) the population of CC1p1h in the sample, affecting indirectly the proportions of the other modes.
}
\end{center}
\end{table}
\begin{table}
\begin{center}
\begin{tabular}{ |c|c|c| } 
\hline
Nominal Model & NRsP 50\%  & NRsP 150\% \\[0cm]
 [cm$^2$] & [cm$^2$] & [cm$^2$] \\ 
\hline
\multicolumn{3}{|c|}{MINERvA}  \\
\hline
$1.94 \times 10^{-39}$ & $2.09 \times 10^{-39}$ & $1.81 \times 10^{-39}$ \\
    & $+7.7\%$ & $-6.7\%$ \\
\hline
\multicolumn{3}{|c|}{T2K} \\
\hline 
$1.46 \times 10^{-39}$ & $1.60 \times 10^{-39}$ & $1.35 \times 10^{-39} $ \\
       & $+9.6\%$ & $-7.5\%$ \\
\hline
\end{tabular}
\caption{\label{Tab:IntegratedCrossSectionNRsP} T2K and MINERvA CC0$\pi$1p total cross-section obtained using NRsP (proton re-scattering probability) configurations of 50\% and 150\%. We use the $d \sigma/ d (\delta\alpha_T)$ differential distribution to perform the numerical integrations. In percentage,  we also show the relative variation with respect to the nominal Model.  }
\end{center}
\end{table}

\section{ Data vs theoretical predictions in terms of the scaling variable} 

\begin{figure}
\begin{center}
\includegraphics[width=0.49\textwidth]{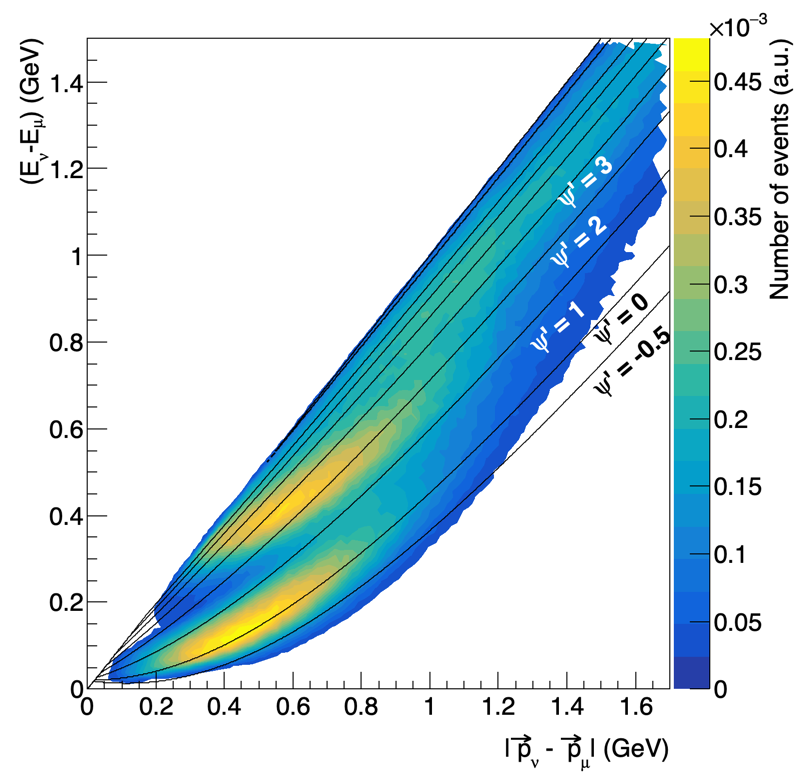}
\includegraphics[width=0.49\textwidth]{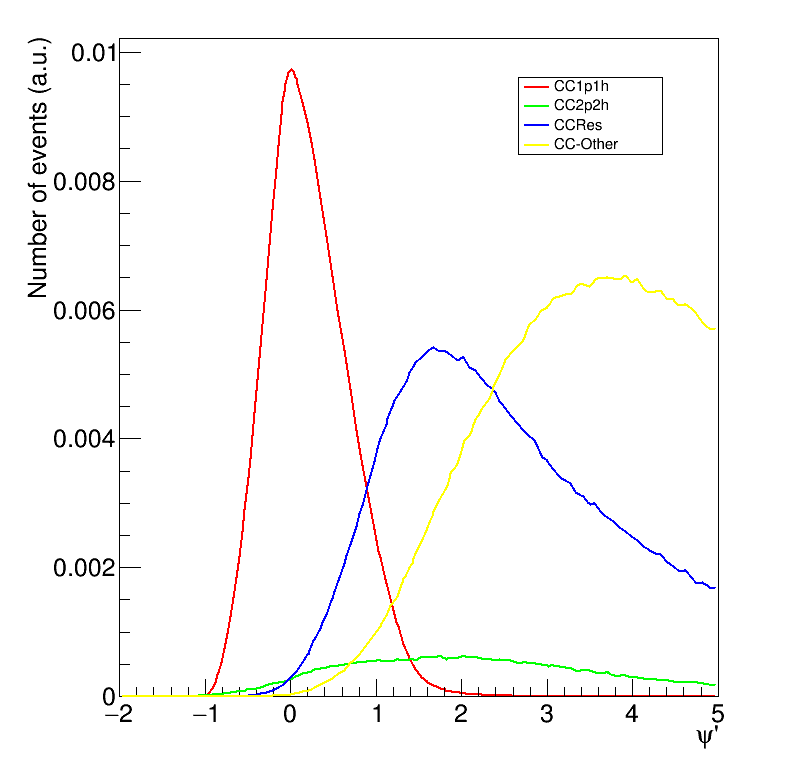}
\caption{ Left: CC inclusive MINERvA [\,$|\vec{q}\,|=|\vec p_{\nu}-\vec p_{\mu}|$, $q^0=(E_{\nu}-E_{\mu})$] 2D  distribution predicted by the NEUT CC inclusive event generator. The black solid lines mark fix $\psi'$ values across the $(q^0, |\vec{q}\,|)-$plane . Right: $\psi'$ distribution of events obtained from the 2D one shown in the left-panel, and  separated by the primary neutrino-nucleon interaction modes.  
\label{Fig:Psi} }
\end{center}
\end{figure}

\begin{figure}
\begin{center}
\includegraphics[width=1.0\textwidth]{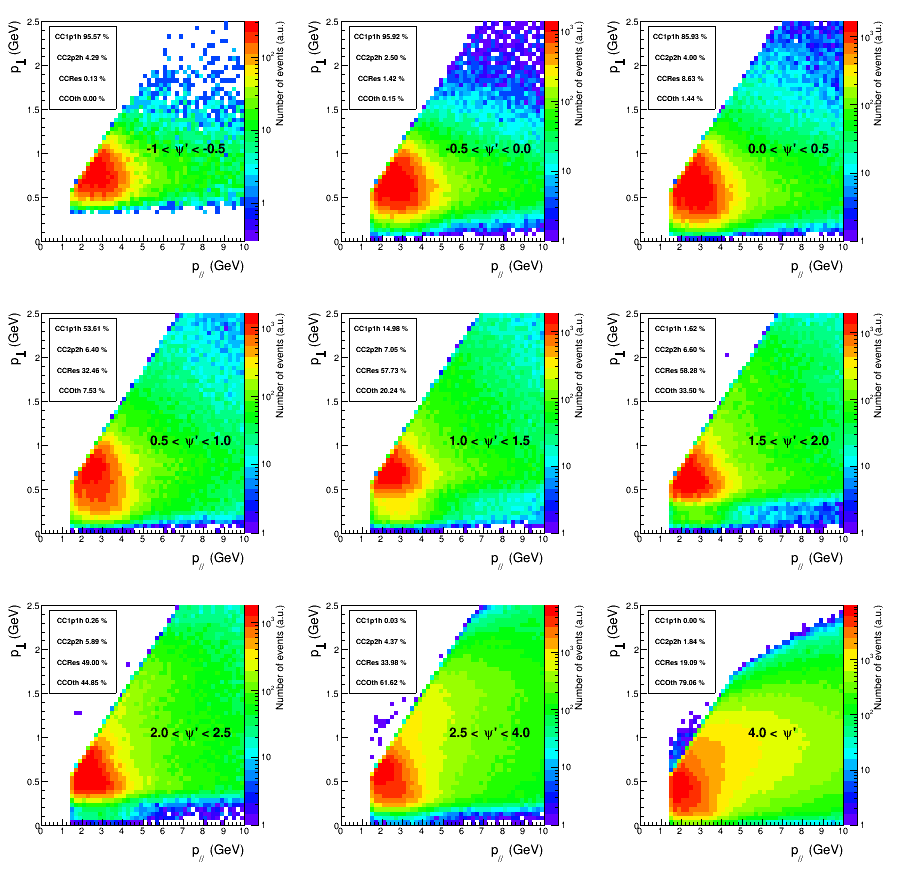}
\caption{ Two dimensional ($p_{\parallel}, p_{\perp}$) lepton distributions accumulated for different $\psi'-$intervals from the NEUT CC inclusive event generator for MINERvA flux. The contributions (in percentage) of the different interaction modes (CC1p1h, C2p2h, CCRes and CCOthers) are also shown in each of the panels.
\label{Fig:fPmuPsi}}
\end{center}
\end{figure}
\begin{figure}
\begin{center}
\includegraphics[width=0.47\textwidth]{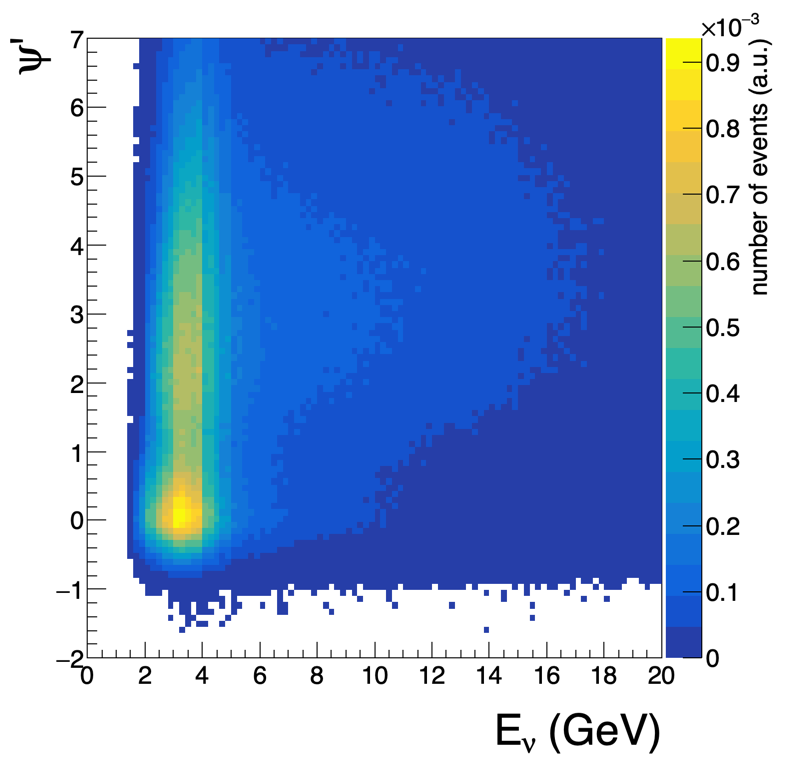}
\includegraphics[width=0.47\textwidth]{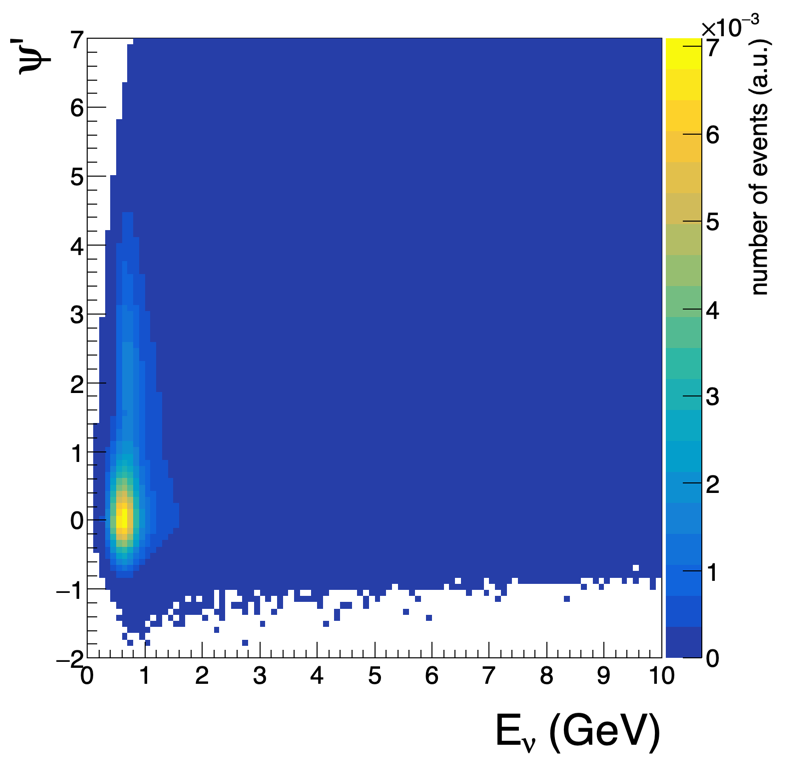}
\includegraphics[width=0.47\textwidth]{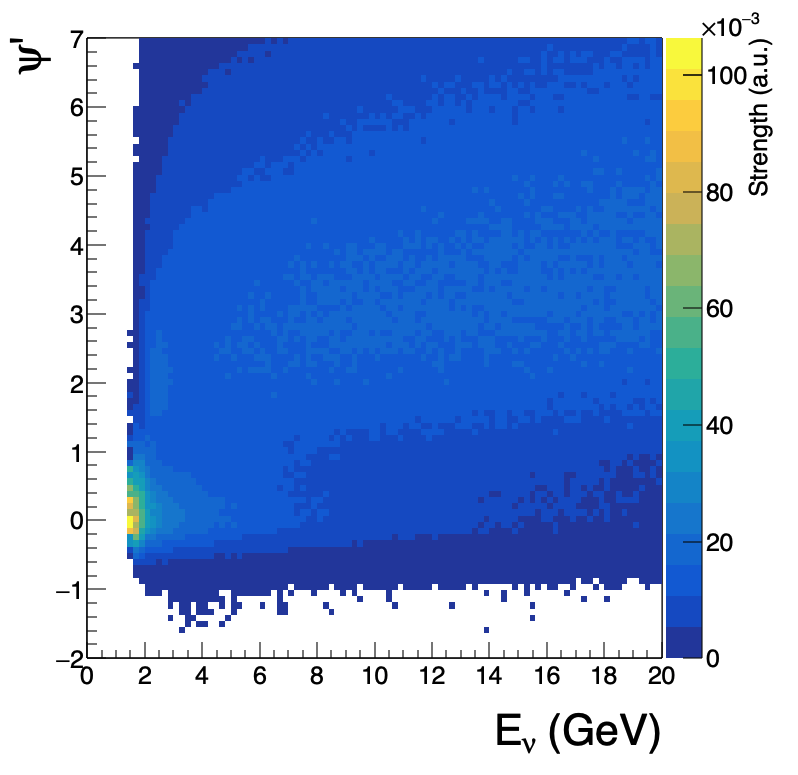}
\includegraphics[width=0.47\textwidth]{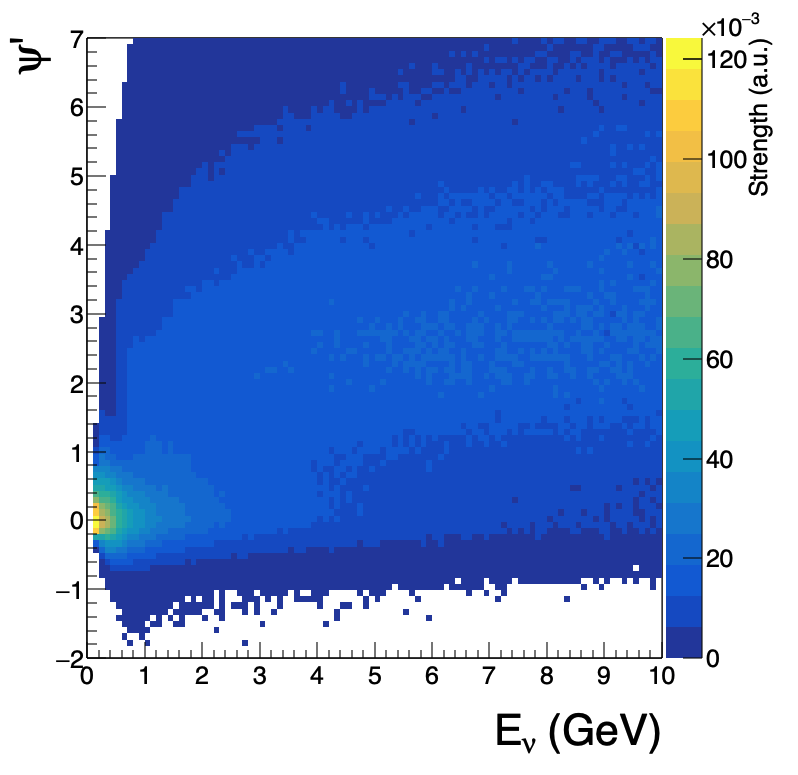}

\caption{ Top: Distribution of the scaling variable $\psi'$ versus the neutrino energy for the CC inclusive MINERvA (left) and T2K (right) experiments, as predicted by NEUT. Bottom: The distributions shown in the upper panels, but normalised to unity by column (energy-bin). 
\label{Fig:PsivsEnu} }
\end{center}
\end{figure}
\begin{figure}
\begin{center}
\includegraphics[width=0.495\textwidth]{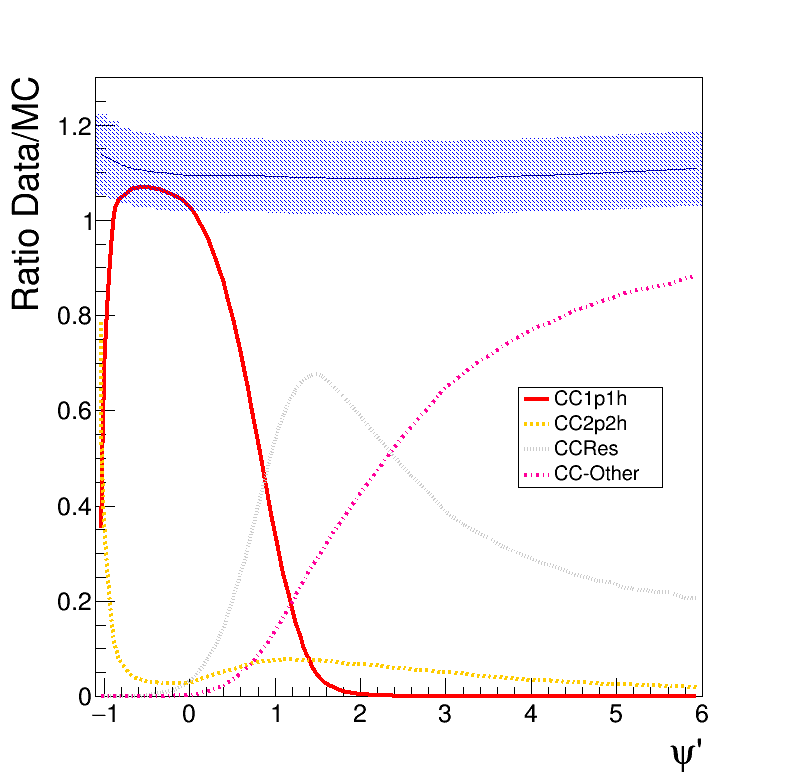}
\includegraphics[width=0.495\textwidth]{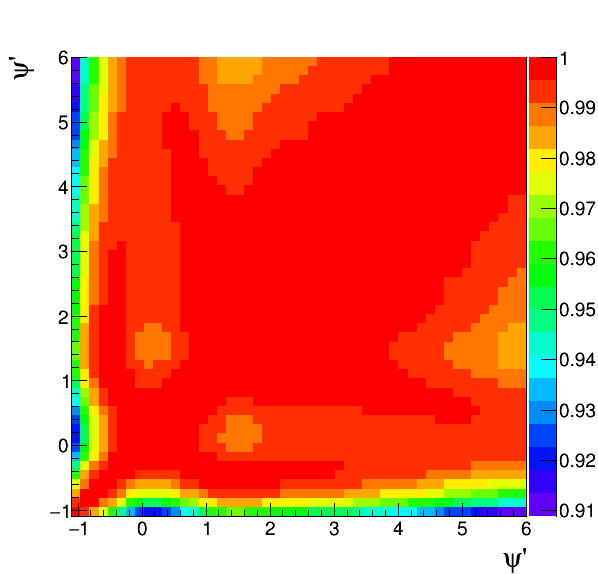}

\caption{ \label{Fig:MinervaCCIncPsi} Left: Data/MC ratio $R(\psi')$ calculated for the MINERvA CC inclusive cross-section, and split into the CC1p1h (red), CC2p2h (yellow), CCRes (gray) and CC-DIS (pink) contributions. Right: $R(\psi')$ correlation matrix computed for the MINERvA CC inclusive cross-sections. }
\end{center}
\end{figure}

\begin{figure}
\begin{center}
\includegraphics[width=0.495\textwidth]{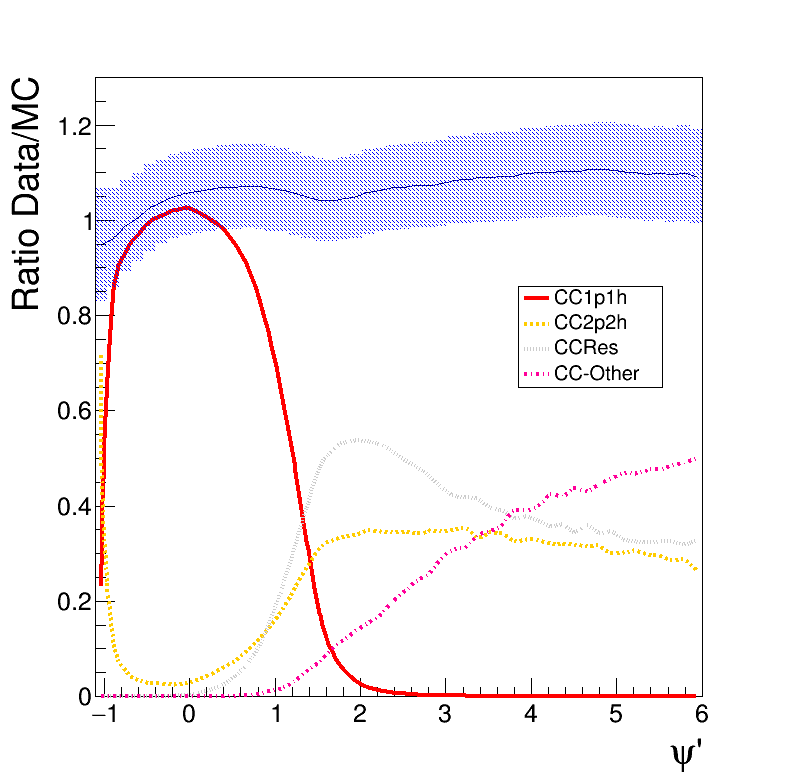}
\includegraphics[width=0.495\textwidth]{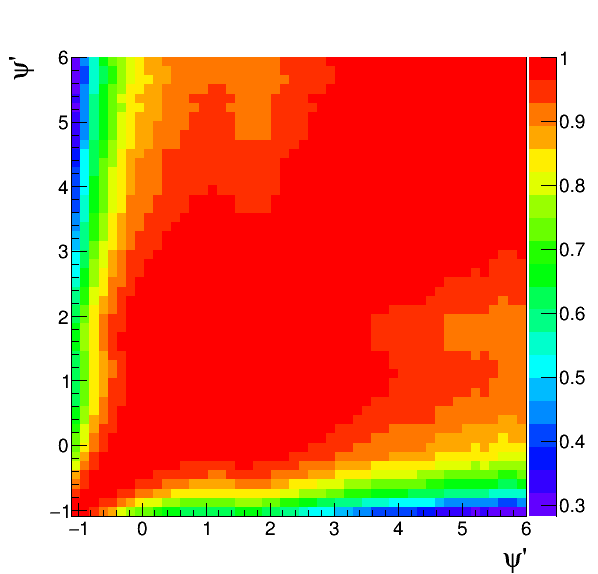}
\caption{ \label{Fig:MinervaCC0piPsi} The same as Fig.~\ref{Fig:MinervaCCIncPsi}, but for the MINERvA CC0$\pi$ cross-section.}
\end{center}
\end{figure}

\begin{figure}
\begin{center}
\includegraphics[width=0.495\textwidth]{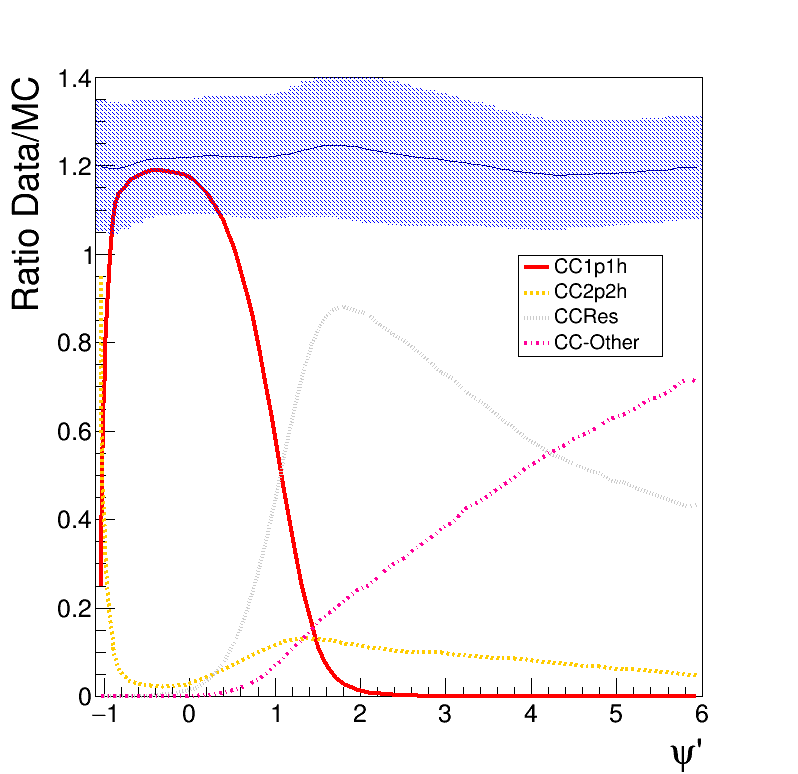}
\includegraphics[width=0.495\textwidth]{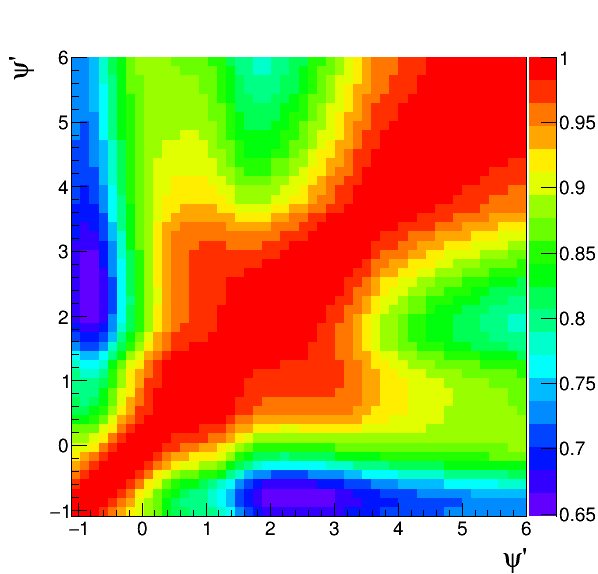}
\caption{\label{Fig:T2KCCIncPsi}  The same as Fig.~\ref{Fig:MinervaCCIncPsi}, but for the T2K CC inclusive cross-section.}
\end{center}
\end{figure}

\begin{figure}
\begin{center}
\includegraphics[width=0.495\textwidth]{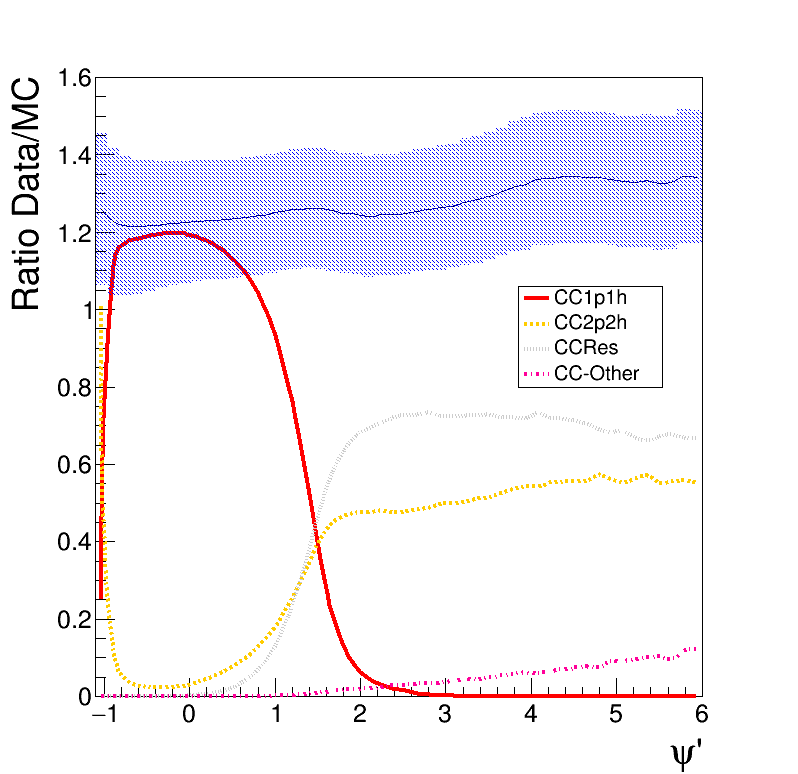}
\includegraphics[width=0.495\textwidth]{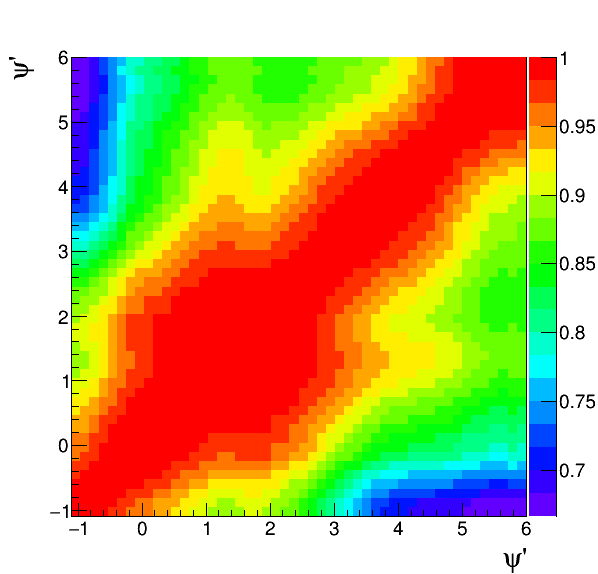}
\caption{\label{Fig:T2KCC0piPsi} The same as Fig.~\ref{Fig:MinervaCCIncPsi}, but for the T2K CC0$\pi$ cross-section. }
\end{center}
\end{figure}

   The scaling variable $\psi'$ has been shown to be a powerful tool to learn details on the neutrino-nucleus interaction~\cite{Gonzalez-Jimenez:2014eqa}. This variable, proposed originally for electron scattering~\cite{Donnelly:1999sw,Barbaro:1998gu}, has been adapted to neutrino-nucleus interactions recently. The scaling variable ($\psi'$) is defined as: 
  \begin{equation}
    \psi' =  \frac{1}{\sqrt{\sqrt{1+\eta_F^2 } - 1}} \frac{ \lambda - \tau }{ \sqrt{(1+\lambda)\tau + \kappa \sqrt{\tau(1+\tau) }} }    
  \end{equation} 
with 
\begin{eqnarray}
\eta_F = \frac{ p_F }{m_{n} } ,\, \kappa = \frac{ |\vec{q}\,| }{ 2 m_{n}},\, \lambda = \frac{ q^0 - E_{\rm shift} }{ 2 m_{n} }\, \text{and},\, \tau = \kappa^2-\lambda^2 
\end{eqnarray}
where, in addition to variables already introduced,  $p_F$ is the Fermi momentum  and $E_{\rm shift}$ is a energy shift, which are  fixed in this discussion to 228~MeV and 20~MeV, respectively.

 The scaling variable $\psi'$ allows to project the energy-momentum  transfer $(q^0, |\vec{q}\,|)$ 2D-sample of events into a 1D distribution, as illustrated  in Fig.~\ref{Fig:Psi} with the NEUT predictions for the CC inclusive MINERvA sample. The full 2D distribution is shown in the left plot, where we also display the $\psi'-$constant curves. We see that a $(q^0, |\vec{q}\,|$) pair determines an unique $\psi'$ value, while the same scaling variable can be constructed out of different energy-momentum  transfer combinations.  The $\psi'-$distribution of events is depicted in the right panel. Though $\psi'$
is based on muon kinematics, the underlying contributions from different mechanisms (1p1h, 2p2h Res and DIS) are
better separated since they lead to different $\psi'-$behaviors, as can be appreciated in the figure. Hence, this variable might provide a  method to effectively disentangle the components of nuclear effects in data and therefore extract valuable constraints on the theoretical model ingredients. This is the objective of the new analysis proposed in this work, and the details of which are discussed below.

The scaling variable $\psi'$ is not accessible to traditional neutrino experiments, since the neutrino energy is not measurable in an event by event basis to compute momentum and energy transfers.  Neutrino experiments usually report their flux-averaged cross-section results as function of muon-kinematics bins\footnote{Experiments normally report $(|\vec{p}_\mu|,\theta_{\mu})$ or $(p_\parallel, p_\perp)$}, $d\sigma_{\rm Exp} / d \vec p_{\mu}$, which can be compared with theoretical  MC differential distributions for the same binning, 
\begin{equation}
  R(\vec p_{\mu})= \frac{d \sigma_{\rm Exp} / d \vec p_{\mu} }{ d \sigma_{\rm MC} / d \vec p_{\mu} } 
\end{equation}
On the other hand within the theoretical MC model, one can associate each event with a value of the scaling variable $\psi'$ (in general, values of the scaling variable comprised in a certain bin, since both neutrino energy and muon-kinematics are binned). We propose to express  the ratio of data to theoretical MC predictions (data/MC, in what follows) as function of $\psi'$. Thus, we define the average  data/MC ratio as:
\begin{eqnarray}
    R(\psi') &=&  \sum_{{\rm events\, MC}\,\in\, \psi' } \, \sum_{ p_{\mu} \theta_{\mu}} f(\vec{p}_\mu| \psi')R(\vec{p}_{\mu}) \label{eq:defR} \\
    f(\vec{p}_\mu| \psi')&=&\frac{N_{\rm MC}(\vec{p}_{\mu}|\psi')}{N(\psi')} \label{eq:defpmu}
\end{eqnarray}
with  $N(\psi')$ the number total of MC generated events, and $N_{\rm MC}(\vec{p}_{\mu}|\psi')$ the number of events with muon kinematics comprised in the bin around $\vec{p}_\mu$, which gives rise to the value $\psi'$ for the scaling variable\footnote{Note that Eq.~\eqref{eq:defR}  admits a matrix interpretation of the type $B = M A$, where $A$ and $B$ are ratio vectors and $M$ is a matrix relating $\psi'$ and the kinematics of the observed muon. }.  Thus, $f(\vec p_{\mu}|\psi')$ is the fraction of events predicted by the theoretical model for a given value of $\psi'$ and muon kinematics $\vec p_{\mu}$, taking into account the neutrino  energy spectrum of the experiment. The distribution $R(\psi')$ is built in such a way that the ratio between data and the theoretical MC results are weighted according to the population $f(\vec p_{\mu}| \psi')$. In the limit in which $f=1$  for one kinematics-bin  and there is no more than one value of $\psi'$ contributing to this bin, $R$ is no more than the ratio of data to MC for this given value of $\psi'$.  This is not the case in most of the experimental bins, but we still expect that some of the deviations from measurements are accumulated in the corresponding value of the scaling variable. 
 
In addition,  the number of events $N_{\rm  MC}(\vec{p}_{\mu}|\psi')$ can be split into the different  mechanisms (1p1h, 2p2h, Res, DIS,...) considered in the theoretical approach implemented in the MC. In Fig.~\ref{Fig:fPmuPsi}, we illustrate the physics content of the 3D $f(\vec{p}_\mu| \psi')$ transfer matrix for the CC inclusive MINERvA sample, as predicted by NEUT. In the figure, we show  the number of events $N_{\rm MC}(\vec{p}_{\mu}|\psi')$, accumulated for different $\psi'$ intervals. As expected from the top panel of Fig.~\ref{Fig:MinervaCCPT}, the largest contributions to $N_{\rm MC}(\vec{p}_{\mu}|\psi')$,  for all $\psi'-$regions, are concentrated in the two dimensional region $[1.5\, {\rm GeV} < p_\parallel < 4 $ GeV] $\times $ $[0.25\, {\rm GeV} < p_\perp < 1 $ GeV]. However, the relative contributions of the different interaction modes (CC1p1h, C2p2h, CCRes and CCOthers) change significantly with the $\psi'-$bin, following a pattern consistent with  the distribution displayed in the right plot of Fig.~\ref{Fig:Psi}.

We stress the trivial observation that a $(p_\parallel,p_\perp)$ pair does not unequivocally determine a  value of $\psi'$, as clearly illustrated in Fig.~\ref{Fig:fPmuPsi}. This is because the neutrino beam is not monochromatic, and for each neutrino energy one has a different relation  between  the $(p_\parallel,p_\perp)$ and $(q^0, |\vec{q}\,|)$ pairs\footnote{One has $q^0 = E_\nu- \sqrt{m_\mu^2+ p_\parallel^2+p_\perp^2}$ and $\vec{q}^{\,2}= (E_\nu-p_\parallel)^2+p_\perp^2$}. As a consequence in a $(p_\parallel,p_\perp)$  distribution, a fix  value of $\psi'$ will not be represented by a curve, but instead by a 2D region, with large overlaps between  different $\psi'-$regions.  We should note, related to this discussion, that the total number of events for a given $\psi'$ [$N(\psi')$ in the definition of $f(\vec{p}_\mu| \psi')$ in Eq.~\eqref{eq:defpmu}] receives sizable contributions from a whole interval of neutrino energies of the incoming broad beam. This is shown in the upper plots of Fig.~\ref{Fig:PsivsEnu} for both MINERvA and T2K CC inclusive samples. We see that for this selection, the latter experiment is much more dominated by the 1p1h reaction mechanism (region of $\psi'$ around one ) than MINERvA, for which resonant and DIS modes, located at higher values of the scaling variable, become more relevant. Moreover, the dispersion of neutrino energies for MINERvA is quite significant, while the high-energy tail for T2K is less important. In the lower plots of Fig.~\ref{Fig:PsivsEnu}, we show the corresponding column (energy-bin) normalized distributions,  where the effect of the neutrino flux times the neutrino cross-section should largely cancel.  We observe  an almost universal pattern, which would be the conditional probability $P(\psi'| E_\nu)$, corrected only by the detector acceptance effects that  are clearly seen at neutrino energies between 0.2~GeV and 3~GeV.

The errors on $R(\psi')$ and the covariance matrix for different values of $\psi'$ can be propagated using the above  definition and the covariance matrix from the experiments. We have applied this algorithm to T2K and MINERvA inclusive and CC0$\pi$ cross-sections.  
 
 \subsection{ Results for $R(\psi')$ }
    
Minerva CC inclusive $R(\psi')$ is shown in the left panel of Fig.~\ref{Fig:MinervaCCIncPsi}, where the contributions of the different reaction channels are also shown. The total distribution shows  a remarkable pattern, with the ratio almost constant, independent of $\psi'$,  around 1.1  (i.e. the MC prediction is some 10\% lower than the data). This means that the proportion of the different reaction channels is well balanced. The 2p2h contribution is very small and it is difficult that this mechanism could significantly influence the overall picture, but both 1$\pi$ and DIS channels  smoothly balance above $\psi' = 1$. We show the $R-$correlation matrix in the right plot of Fig.~\ref{Fig:MinervaCCIncPsi}. The correlation is larger than 90\% for any pair of of $\psi'$ values. This is a consequence of both the initial correlation between experimental results and the fact that several $\psi'-$values  contributes to the same muon-kinematics bin.

We repeat in Fig.~\ref{Fig:MinervaCC0piPsi} the exercise with the  CC0$\pi$ MINERvA data. The tendency observed is similar to that seen for the CC inclusive. MC predicts smaller cross-section by a similar amount (10\%), though in the CC0$\pi$ case some structure is observed with  a small deficit of MC below $\psi'=0$. On the contrary, the region above $\psi'= 2$ shows an excess on the MC predictions with respect to data. This region is dominated by 2p2h and $1\pi$ (Res) channels. It is important to notice that these results do not call for a large modification of the 2p2h contribution as requested by the calorimetric measurements in MINERvA~\cite{Rodrigues:2015hik}. The bin to bin correlation is shown in the right panel of the figure.  The observed correlations are smaller than in the previous case, with some regions below 70\%. This might explain the appearance of some structure in the CC0$\pi$ ratio plot.

\begin{figure}
\begin{center}
\includegraphics[width=0.49\textwidth]{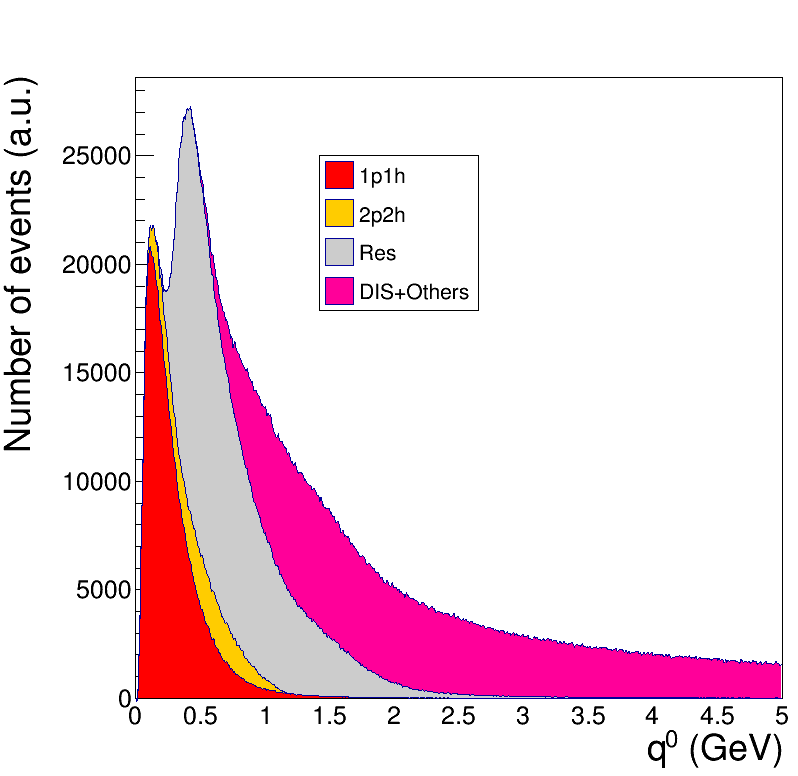}
\includegraphics[width=0.49\textwidth]{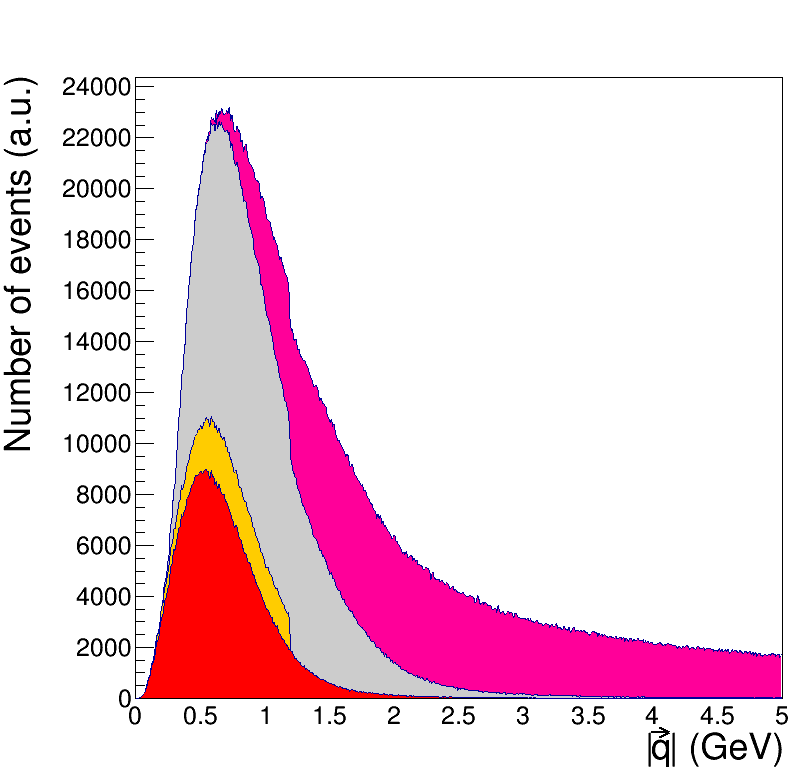}
\includegraphics[width=0.49\textwidth]{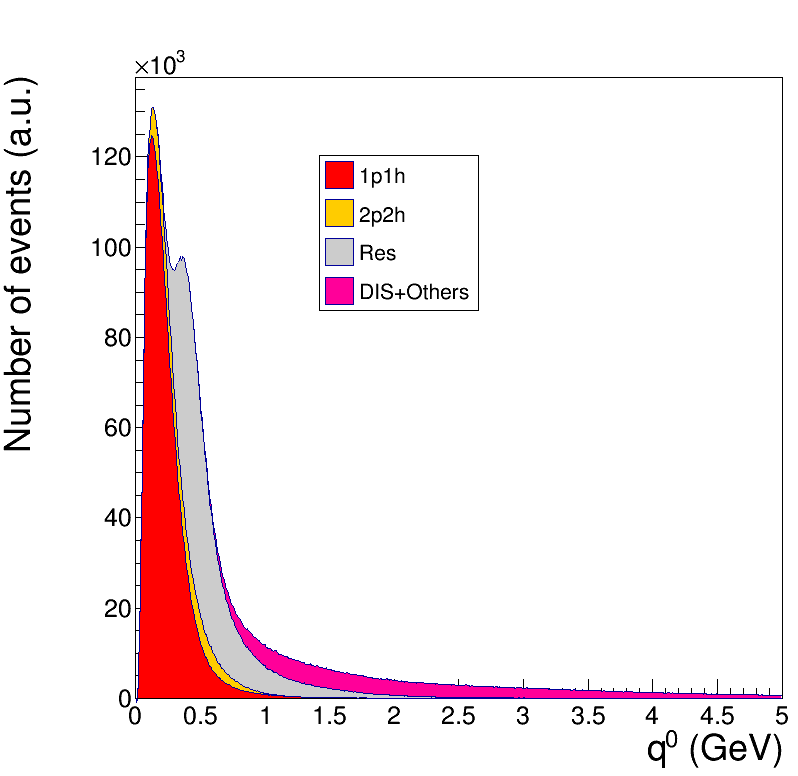}
\includegraphics[width=0.49\textwidth]{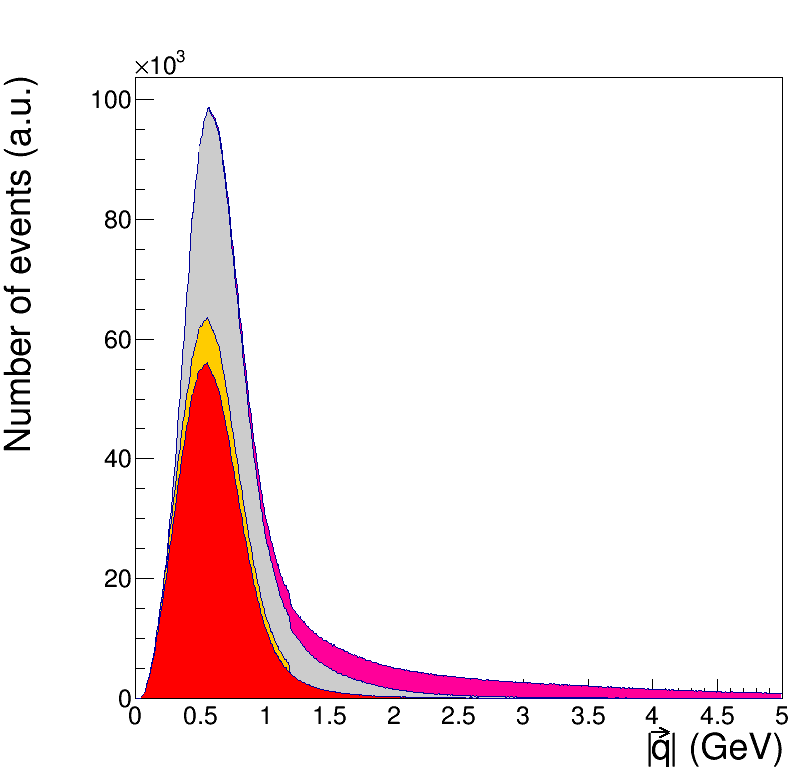}

\caption{\label{Fig:Q0Q2MinervaT2K} MINERvA (up) and T2K (bottom) energy $q^0$ (left) and momentum $|\vec{q}\,|$ (right) transfer distributions  for the CC inclusive selections. }
\end{center}
\end{figure}
\begin{figure}
\begin{center}
\includegraphics[width=0.49\textwidth]{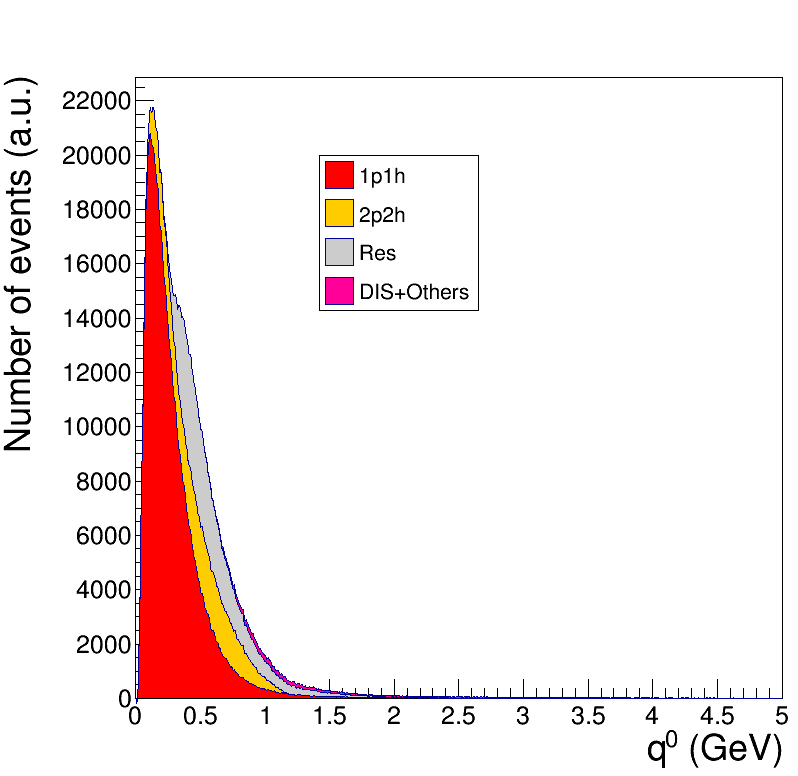}
\includegraphics[width=0.49\textwidth]{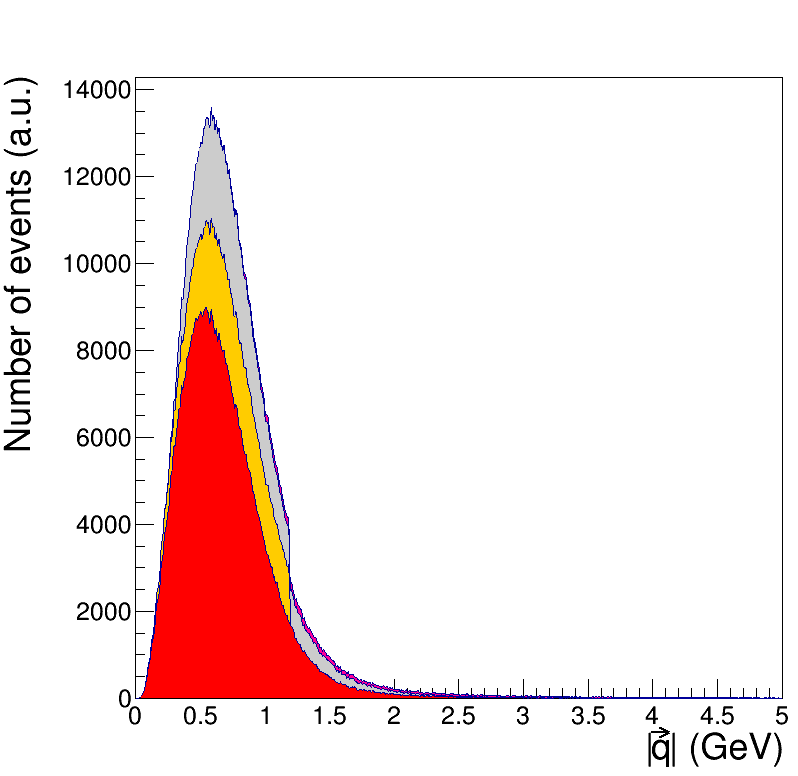}
\includegraphics[width=0.49\textwidth]{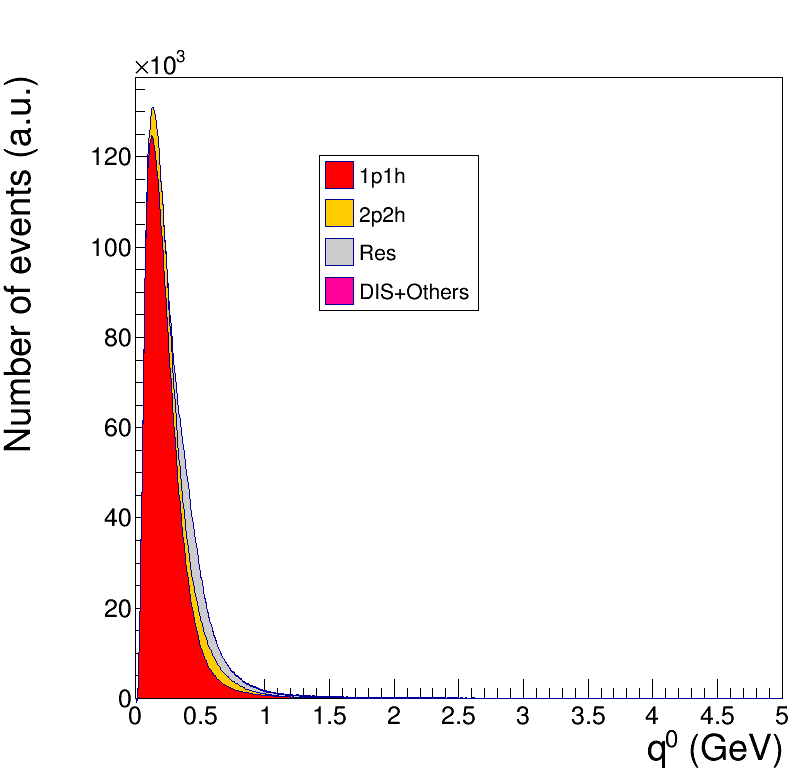}
\includegraphics[width=0.49\textwidth]{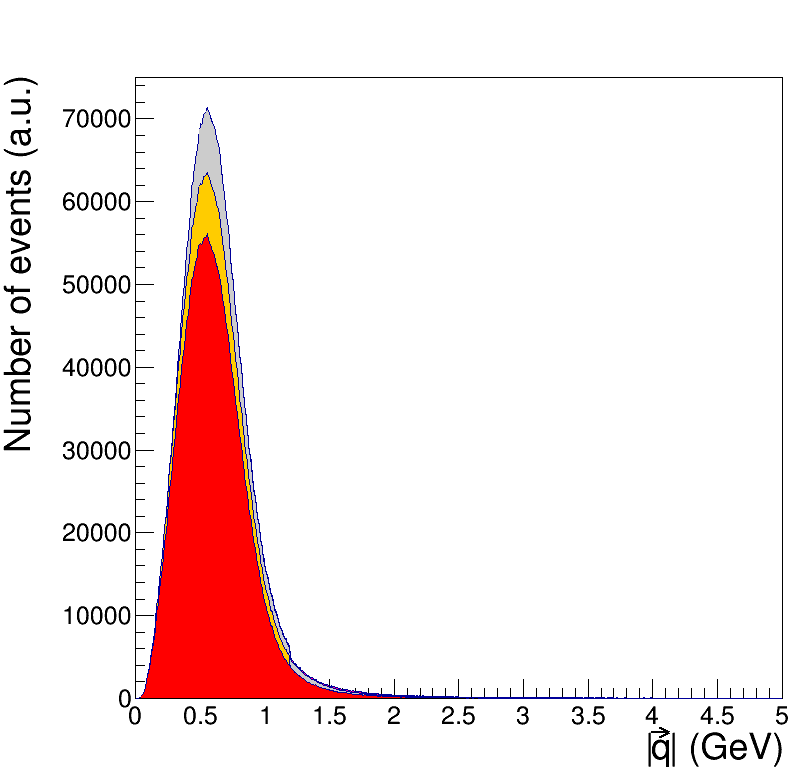}
\caption{\label{Fig:Q0Q2MinervaT2KCC0pi} The same as Fig.~\ref{Fig:Q0Q2MinervaT2K} for CC0$\pi$ selections.   }
\end{center}
\end{figure}
\begin{figure}
\begin{center}
\includegraphics[width=0.49\textwidth]{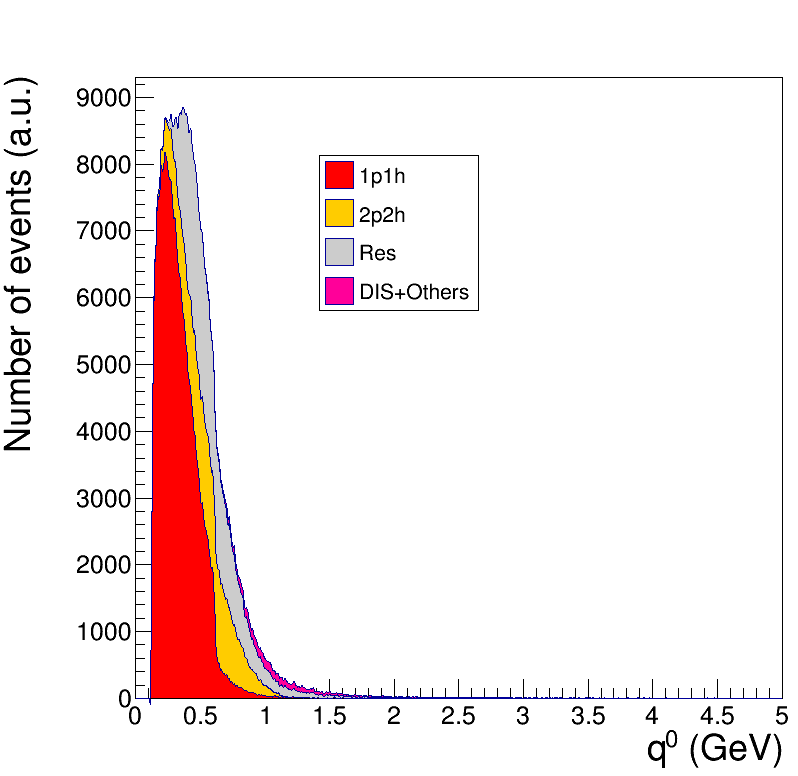}
\includegraphics[width=0.49\textwidth]{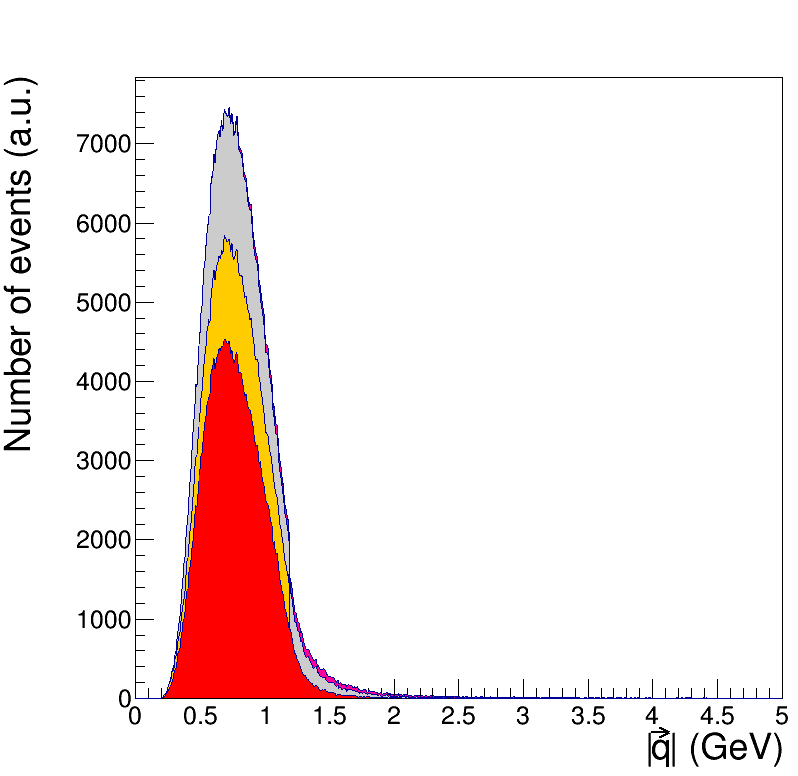}
\includegraphics[width=0.49\textwidth]{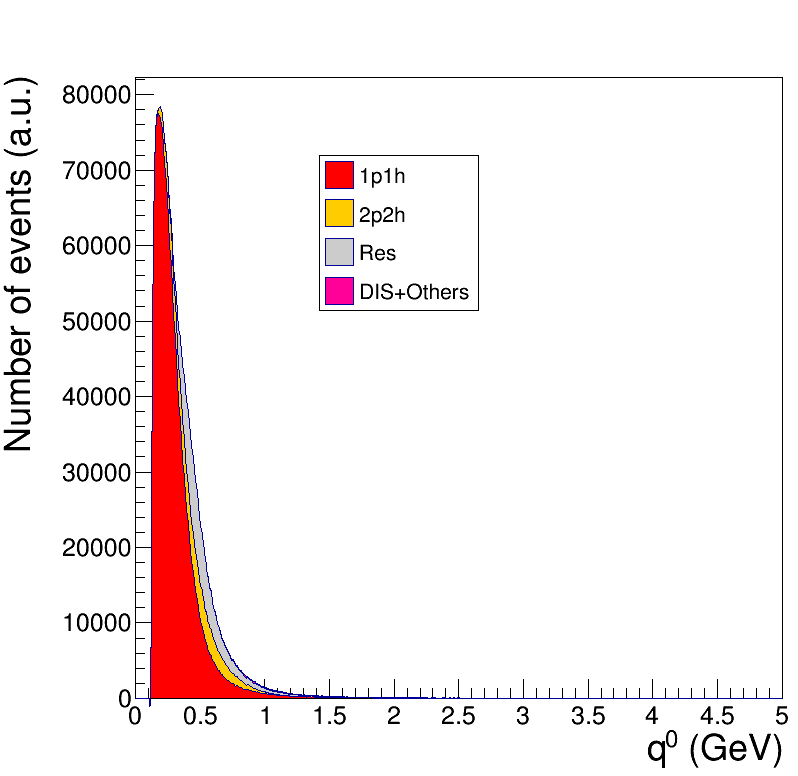}
\includegraphics[width=0.49\textwidth]{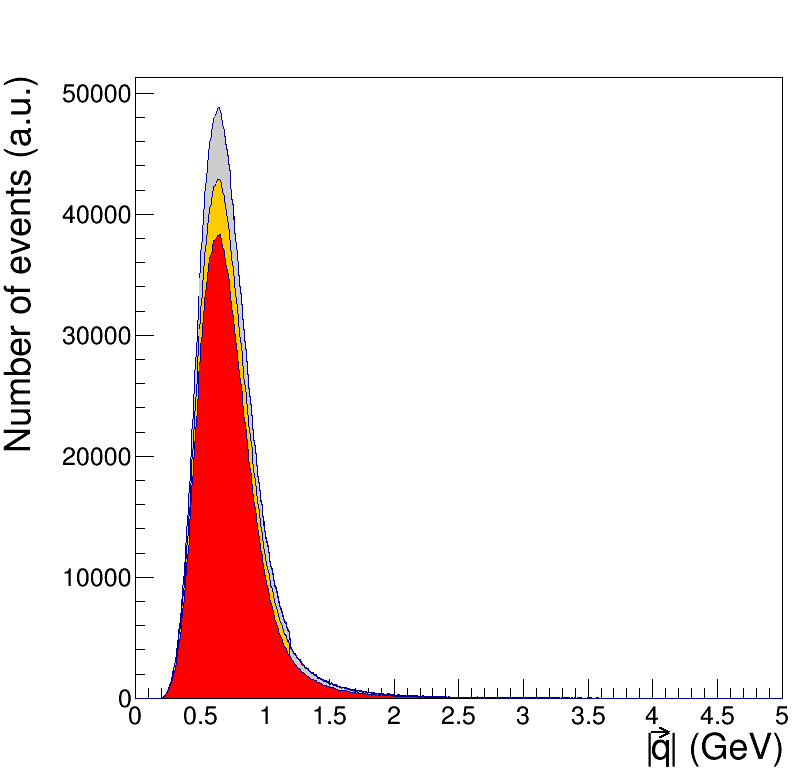}
\caption{\label{Fig:Q0Q2MinervaT2KTransv}  The same as Fig.~\ref{Fig:Q0Q2MinervaT2K} for CC0$\pi$1p selections.   }
\end{center}
\end{figure}

The $R(\psi')$ results for T2K  CC inclusive  and CC$0\pi$ samples are shown in Figs.~\ref{Fig:T2KCCIncPsi} and \ref{Fig:T2KCC0piPsi}, respectively. The observed patterns  in both cases are similar to those discussed above for the MINERvA experiment. Our model predicts smaller cross-section than the measurements  by about 20\% ($R(\psi')\sim 1.2$), while the distribution of the ratio as function of $\psi'$ is again rather flat. There is some reflection of the CCRes contribution in Fig.~\ref{Fig:T2KCCIncPsi} around $\psi'\sim 1.5$, which might be an indication of an even smaller single pion cross-section. Nevertheless and despite the large correlations, the evidence is weak, given the flat behavior of the ratio and the large errors.  Similar low cross-section predictions were also observed in the T2K CC1$\pi$ results~\cite{Abe:2019arf}. The impact of the possible CC1$\pi$ miss-modelling is minimised in the case of CC0$\pi$ where the contribution of CC1$\pi$ becomes as the same level than CC 2p2h, see Table~\ref{Tab:SampleComposition}. The region around the CCQE peak shows a flat dependency with the scaling variable contrary to the MINERvA CC0$\pi$ results which exhibit a visible decrease below $\psi'< 0$. It is also observed that the prediction is balanced between the different reaction channels. The correlation matrices are shown in the right panels of Figs.~\ref{Fig:T2KCCIncPsi} and \ref{Fig:T2KCC0piPsi}. The correlations for both CC inclusive and CC$0\pi$ are smaller than in the case of MINERvA, 60\% correlation between values of $\psi'$ below 0 and higher than 1.5. This reduced correlation gives more credibility to the tendencies of $R(\psi')$.  

The large errors ($\approx 10\%$ for T2K and $\approx 5\%$ for MINERvA) are the consequence of large positive correlations enhancing the experimental errors. 
     
To understand the large correlations across all values of $\psi'$ both for T2K and MINERvA, we estimated the correlation removing all the off-diagonal terms in the covariance matrices provided by the experiment. In this case, the ratios $R(\psi')$ for different values of the scaling variable can be correlated only because receive contributions from the same muon-kinematics bin for different neutrino energies,  contained in the non-monochromatic beam (see the discussion above of Figs.~\ref{Fig:fPmuPsi} and ~\ref{Fig:PsivsEnu}).  The smallest observed correlation is reduced from 90\% to 40\% in the case of MINERvA and from 60\% to 5\% in the case of T2K. The better figure in the CC0$\pi$ T2K sample  might be a consequence of the narrow neutrino beam at T2K, which  allows to separate better the regions dominated by 1p1h and the resonant and DIS components. The experimental correlations errors come mostly from flux uncertainties, statistical and systematic experimental errors  but also from bin to bin migrations in the extraction of the cross-section. Improvement in flux determination, larger statistics to reduce the bin size and select the proper data representation might improve the conclusions of this study.

\section{Conclusions } 

 We have presented an exclusive final state model to describe CC1p1h interactions. The approach is based on a LFG picture of the nucleus and uses a consistent implementation of the removal energy, that provides an estimation of the excitation of the final nuclear system. The model has been included in NEUT to profit from the existing simulation of 2p2h, pion production and DIS mechanisms  and on the transport simulation of the hadrons inside the nucleus after the interaction. Predictions are simultaneously compared to the most recent T2K and MINERvA inclusive, CC$0\pi$ and TKI variable results, showing an acceptable agreement with the data from both experiments.  Results from T2K suffer from low statistics, but they also show worse agreement with the model predictions. This might be an indication of some energy dependency that is not properly accounted by this implementation. The correct modelling of the energy removal reduces the amount of interactions at low $q^2$ and the total CC1p1h cross-section facilitating the agreement with the experimental results.  On the other hand, the overall good description of the MINERvA TKI variables found here and, in general of its CC0$\pi$ data-sample, does not support a large modification of the 2p2h contribution as requested by the calorimetric measurements of that collaboration~\cite{Rodrigues:2015hik}. The latter conclusion agrees with the findings in Ref.~\cite{Lu:2018stk} with NuWro~\cite{Golan:2012wx} and GiBUU~\cite{Mosel:2019vhx} event generators (GiBUU results can be found in the Supplemental Material  for Ref.~\cite{Coplowe:2020yea}). However, we should point out that the re-weight of the 2p2h strength proposed in  \cite{Rodrigues:2015hik} is based on the inclusive sample, which
is a superset of the CC0$\pi$ and CC0$\pi$1p data-sets considered in this work. 
 
We have also proposed a novel comparison between flux-folded data and MC theoretical predictions accumulated in bins of the scaling variable $\Psi'$, which can be used to signal possible  deficiencies of theoretical schemes.

A microscopic interpretation of the relevant reaction mechanisms becomes essential in neutrino oscillation experiments
in order to achieve a correct reconstruction of the incoming neutrino kinematics, free of conceptual biasses.
Studies, as the one presented in this work, are of the utmost importance for the ambitious experimental program which is
underway to precisely determine neutrino properties, test the three-generation paradigm,
establish the order of mass eigenstates and investigate leptonic CP violation.

\appendix

\section{Appendix: Comparison of momentum and energy transfer distributions }
\label{App:Momentum transfer}

 The different experimental selection criteria might bias the momentum and energy transfers. The comparison of the different accessible  $(q^0,|\vec{q}\,|)-$phase space provide some indications of possible deficits in the models. 
 
 The model estimation for the energy and momentum transfer distributions for the MINERvA and T2K CC inclusive event selections are shown in Fig.~\ref{Fig:Q0Q2MinervaT2K}. The cut off in $|\vec{q}\, |$ implemented in the CC2p2h model is clearly visible in the MINERvA distributions. The CC2p2h cross-section at the cutoff $|\vec{q}\, |=$ 1.3 GeV represents around 10\% of the total one for this momentum transfer. Although, we find similar drops for MINERvA and T2K, there are  more contributions to the cross-section in the first experiment above $|\vec{q}\,|=$ 1.3 GeV, and hence we  expect that the implementation of this cutoff to make a bigger impact in the total cross-section determination for MINERvA.  As expected, the resonant, 2p2h and DIS contributions are larger for the MINERvA energies, see Table \ref{Tab:SampleComposition}. 
 
\begin{figure}
\begin{center}
\includegraphics[width=\textwidth]{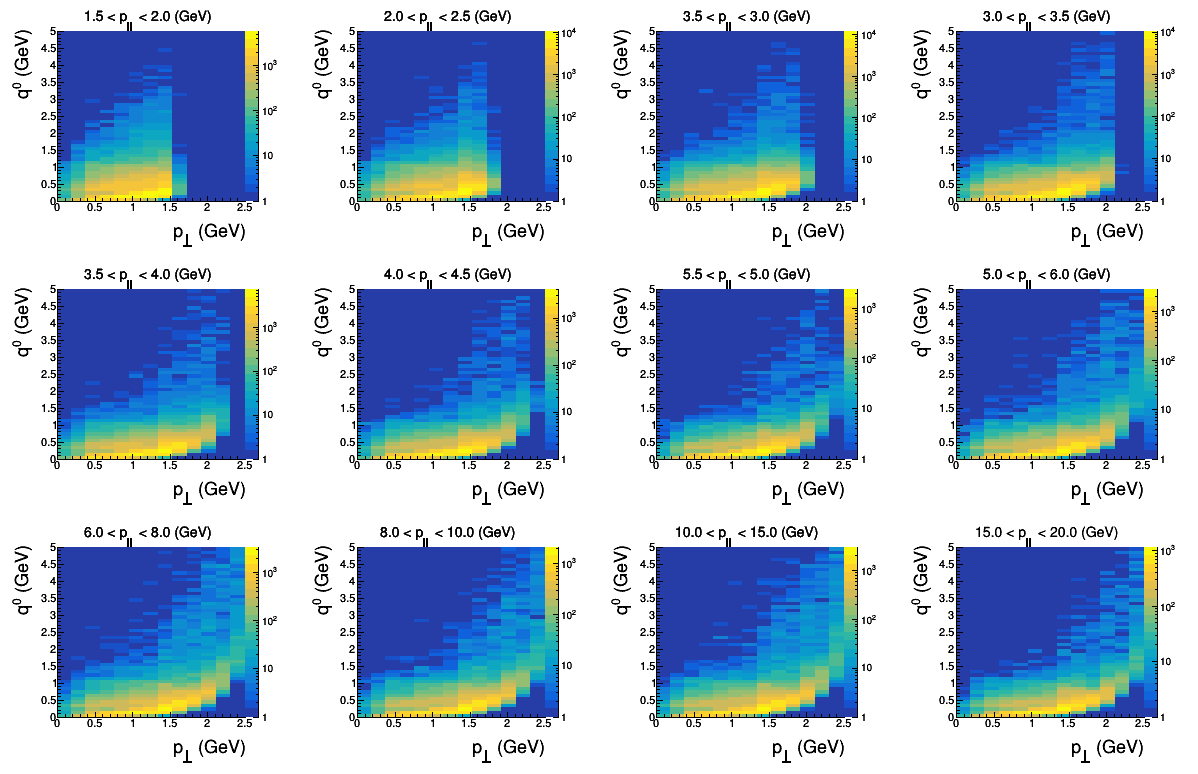}

\caption{\label{Fig:Q0vsPT} $q^0$ vs $p_{\perp}$ for MINERvA CC0$\pi$ sample  for the same $p_\parallel$ bins as in Fig.~\ref{Fig:MinervaCCPT}. The color code represents the number of events in arbitrary units. }
\end{center}
\end{figure}

\begin{figure}
\begin{center}
\includegraphics[width=\textwidth]{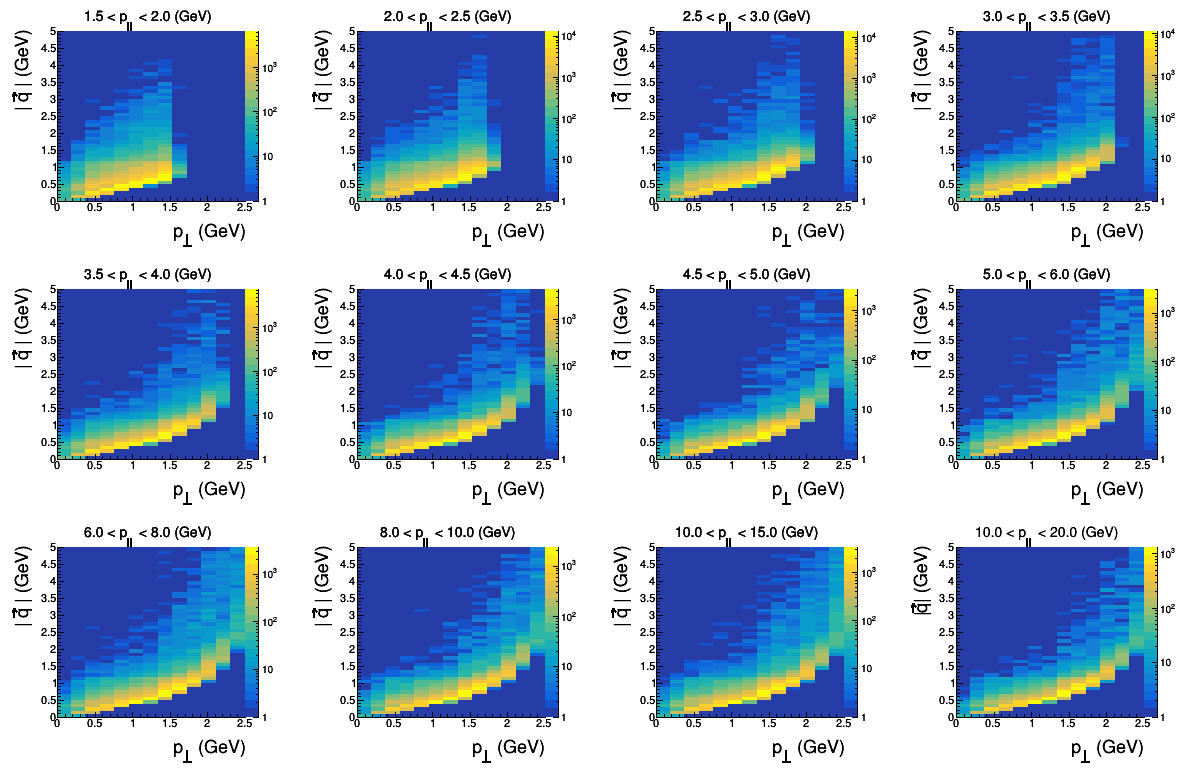}
\caption{\label{Fig:Q3vsPT} $|\vec{q}\,|$ vs $p_{\perp}$ for MINERvA CC0$\pi$ sample for the same $p_\parallel$ bins as in Fig.~\ref{Fig:MinervaCCPT}. The color code represents the number of events in arbitrary units. }
\end{center}
\end{figure}

 The predicted energy and momentum transfer distributions for the MINERvA and T2K CC0$\pi$ event-selections are shown next in Fig.~\ref{Fig:Q0Q2MinervaT2KCC0pi}. As in the previous case, the cut off $|\vec{q}\, | \le $ 1.3 GeV  for the CC2p2h contribution is clearly visible, both for MINERvA and T2K distributions, with the drop at the cutoff amounting around  10\%  of the total cross-section   As expected, the CC0$\pi$ event-sample has smaller contamination from DIS and resonant processes, which leads to a small bias in the event selection.  Moreover, the  CC1p1h contributions in this data selection show a dependence on $q^0$ and $|\vec{q}\,|$ similar as  that observed for the CC inclusive ones.

 The $q^0$ and $|\vec{q}\,|$ event distributions used for the TKI variable analysis (CC0$\pi$1p) are shown in Fig.~\ref{Fig:Q0Q2MinervaT2KTransv}. The significant reduction of the CCRes and CCOther contributions is evident in the plots  and it can be also seen in Table~\ref{Tab:SampleComposition}. The detector acceptance cut employed for the MINERvA  CC0$\pi$1p selection of events is observed as a change of slope  around $q^0=0.6$ GeV, clearly visible  in the 1p1h distribution. In the case of T2K, the hard cutoff is not visible and the distributions of momentum and energy transfers are narrower and shifted towards lower values, with smaller contamination from Res and DIS-Others components (see also Table \ref{Tab:SampleComposition}).  In Fig.~\ref{Fig:Q0Q2MinervaT2KTransv}, one can also observe a shift in the value of $|\vec{q}\, |$. Actually, the mean value of $|\vec{q}\, |$ in the MINERvA (T2K) 1p1h component  moves from 0.69~GeV (0.61~GeV) in the  CC0$\pi$ data-sample to 0.77~GeV (0.73~GeV) in the CC0$\pi$1p one. This is a consequence of requesting a proton above 0.45~GeV in the detector. 

Finally in Figs.~\ref{Fig:Q0vsPT} and \ref{Fig:Q3vsPT}, we show for the MINERvA CC$0\pi$ sample, the $q^0$ and $|\vec{q}\,|$ distributions as a function of $p_\perp$ for the same $p_\parallel$ binning as in Fig.~\ref{Fig:MinervaCCPT}.  A high occupancy region corresponding to the CC1p1h contribution is clearly observed. A second one, mainly due to  CC2p2h and CCRes events, is also visible in $q^0$, see Fig.~\ref{Fig:Q0vsPT} (note the logarithm scale in the $z-$coordinate). This second enhanced region is less visible in $|\vec{q}\,|$, see Fig.~\ref{Fig:Q3vsPT}, except for the low longitudinal momentum bins. The distributions also show that $|\vec{q}\,|$ values are similar for a given $p_\perp$ independently of the longitudinal momentum except for the case with $p_{\parallel} < 3$ GeV. The main differences between data and MC observed in Fig.~\ref{Fig:MinervaCCPT} as function of $p_{\parallel}$ are most probably caused by the CCRes and CC2p2h large $q^0$ contributions.

\acknowledgments

This work was supported by the Swiss National Foundation Grant No. 200021\_85012, the Spanish Ministerio de Econom\'{i}a y Competitividad and the European Regional Development Fund under contract FIS2017-84038-C2-1-P, the EU STRONG-2020 project under the program H2020-INFRAIA-2018-1, grant agreement no. 824093 and by  Generalitat Valenciana under contract PROMETEO/2020/023. We thank the MINERVA collaboration for letting us notice the publication of their new results.


\bibliographystyle{unsrt}

\bibliography{Bibliography}

\end{document}